\documentclass[times,resetfootnote,twocolumn,tighten]{aastex7}

\usepackage{amsmath}
\usepackage{xspace}                       % for appropriate spacing after \mircx
\newcommand{\mircx}{\mbox{MIRC-X}\xspace} % Prevent line breaking on MIRC-X hyphen
\newcommand{\hband}{\mbox{H-band}\xspace}
\newcommand{\kband}{\mbox{K-band}\xspace}
\usepackage{multirow}
\usepackage{afterpage}

\usepackage{savesym}
\savesymbol{tablenum}
\usepackage[separate-uncertainty=true,multi-part-units=single]{siunitx}
\restoresymbol{SIX}{tablenum}

\DeclareSIUnit \inch {\!\ensuremath{^{\prime\prime}}}
\DeclareSIUnit \bar {bar}
\DeclareSIUnit \parsec {pc}
\DeclareSIUnit \solarmass {\mbox{M$_\odot$}}
\DeclareSIUnit \solarluminosity {\mbox{L$_\odot$}}
\DeclareSIUnit \solarradius {\mbox{R$_\odot$}}
\DeclareSIUnit \milliarcsecond {mas}
\DeclareSIUnit \au {au}
\DeclareSIUnit \year {yr}

\shorttitle{HD~163296 Inner Disk}
\shortauthors{Setterholm et al.}
\submitjournal{\aj}
\received{2024 July 28}
\revised{2024 December 20}
\revised{2025 March 25}
\accepted{2025 April 14}

\begin{document}

\title{The Dynamic Inner Disk of a Planet Forming Star}

% \correspondingauthor{Benjamin Setterohlm}
% \email{setterholm@mpia.de}

\author[orcid=0000-0001-5980-0246]{Benjamin R. Setterholm}
\affiliation{Department of Astronomy, University of Michigan; Ann Arbor, MI, USA}
\affiliation{Max-Planck-Institut für Astronomie; Heidelberg, Germany}
\email{setterholm@mpia.de}

\author[orcid=0000-0002-3380-3307]{John D. Monnier}
\affiliation{Department of Astronomy, University of Michigan; Ann Arbor, MI, USA}
\email{monnier@umich.edu}

\author[orcid=0000-0002-8376-8941]{Fabien Baron}
\affiliation{Department of Physics and Astronomy, Georgia State University; Atlanta, GA, USA}
\email{fbaron@gsu.edu}

\author[orcid=0000-0001-7258-770X]{Jaehan Bae}
\affiliation{Department of Astronomy, University of Florida; Gainesville, FL, USA}
\email{jbae@ufl.edu}

\author[orcid=0000-0002-9491-393X]{Jacques Kluska}
\affiliation{Institute of Astronomy, KU Leuven; Leuven, Belgium}
\affiliation{Schneider Electric; Grenoble, France}
\email{jacques.kluska@protonmail.com}

\author[orcid=0000-0001-6017-8773]{Stefan Kraus}
\affiliation{Department of Physics and Astronomy, University of Exeter; Exeter, UK}
\email{S.Kraus@exeter.ac.uk}

\author[orcid=0000-0002-3950-5386]{Nuria Calvet}
\affiliation{Department of Astronomy, University of Michigan; Ann Arbor, MI, USA}
\email{ncalvet@umich.edu}

\author[orcid=0000-0002-1788-9366]{Nour Ibrahim}
\affiliation{Department of Astronomy, University of Michigan; Ann Arbor, MI, USA}
\email{inoura@umich.edu}

\author[orcid=0000-0002-1779-8181]{Evan Rich}
\affiliation{Department of Physics and Astronomy, University of Nebraska; Lincoln, NE, USA}
\email{erich3@unl.edu}

\author[orcid=0000-0002-2208-6541]{Narsireddy Anugu}
\affiliation{The CHARA Array of Georgia State University; Mount Wilson, CA, USA}
\email{nanugu@chara-array.org}

\author[orcid=0000-0001-9764-2357]{Claire L. Davies}
\affiliation{Department of Physics and Astronomy, University of Exeter; Exeter, UK}
\email{C.Davies3@exeter.ac.uk}

\author[orcid=0000-0002-1575-4310]{Jacob Ennis}
\affiliation{Department of Astronomy, University of Michigan; Ann Arbor, MI, USA}
\email{ennisj@umich.edu}

\author[orcid=0000-0002-3003-3183]{Tyler Gardner}
\affiliation{Department of Physics and Astronomy, University of Exeter; Exeter, UK}
\email{T.B.Gardner@exeter.ac.uk}

\author[orcid=0000-0001-8837-7045]{Aaron Labdon}
\affiliation{European Southern Observatory; Santiago, Chile}
\email{Aaron.Labdon@eso.org}

\author[orcid=0000-0001-9745-5834]{Cyprien Lanthermann}
\affiliation{The CHARA Array of Georgia State University; Mount Wilson, CA, USA}
\email{cyprien@chara-array.org}

\author[orcid=0000-0001-5415-9189]{Gail Schaefer}
\affiliation{The CHARA Array of Georgia State University; Mount Wilson, CA, USA}
\email{schaefer@chara-array.org}

\begin{abstract}
    Planets are a natural byproduct of the stellar formation process, resulting from local aggregations of material within the disks surrounding young stars. Whereas signatures of gas-giant planets at large orbital separations have been observed and successfully modeled within protoplanetary disks, the formation pathways of planets within their host star's future habitable zones remain poorly understood. Analyzing multiple nights of observations conducted over a short, two-month span with the \mircx and PIONIER instruments at the CHARA Array and VLTI, respectively, we uncover a highly active environment at the inner-edge of the planet formation region in the disk of HD~163296. In particular, we localize and track the motion of a disk feature near the dust-sublimation radius with a pattern speed of less than half the local Keplerian velocity, providing a potential glimpse at the planet formation process in action within the inner astronomical unit. We emphasize that this result is at the edge of what is currently possible with available optical interferometric techniques and behooves confirmation with a temporally dense followup observing campaign.
\end{abstract}


\section{Introduction}

Pre-main-sequence, young stellar objects (YSOs) are surrounded by circumstellar disks composed of gas and dust, typically extending out several hundred \unit{\au}. Planets coalesce within these disks, though the formation pathways and typical timescales involved are not yet completely understood. The success of exoplanet surveys in the past 30 years has revealed a large variety of planetary system architectures, suggesting that the formation process is highly dynamic and stochastic \citep{Zhu+2021}. A notable recurrence observed in many exoplanet systems are the so-called ``hot-Jupiters'' -- massive planets orbiting well within an \unit{\au} of the central star. The preponderance of such worlds has challenged traditional ideas of the planetary formation process \citep{Dawson+2018}; it is still unclear whether they develop primarily at larger orbits and later migrate inward or whether they aggregate their mass largely in-situ.

The configuration of the protoplanetary disks themselves in the vicinity of the central star, where the latter heats its immediate surroundings above the sublimation temperature of typically assumed astrophysical (silicate) dust species, is also not yet well understood \citep{Dullemond+2010}. The geometry of the interface between the gaseous region and the proper start of the dusty disk is important both for deciphering the accretion physics of material onto the star as well as providing a boundary limiting the innermost orbital distances of eventual planets. The dust sublimation rim is also thought to be the dominant contributor to the near-infrared radiation, exemplified by the famous size-luminosity relation \citep{Monnier+2002}. In many YSO disk systems, near-infrared flux variability (sometimes in excess of \qty{10}{\percent}) is observed on timescales of days to weeks \citep{Carpenter+2001,Flaherty+2012}, suggesting that the zone near the sublimation rim can be highly dynamic. Some systems have also been seen to exhibit shadows cast on the outer disk moving at speeds only explainable by activity within the inner disk \citep{Benisty+2023}. Such observations have long been speculated to be caused by planetary activity at close orbits \citep{Espaillat+2011}.

While much progress has been made in understanding the macroscopic physics occurring in protoplanetary disks through hydrodynamic simulation, physical models still need to be tested and tuned with spatially-resolved observational data. Current and next-generation ``conventional'' observatories fundamentally lack the sub-milliarcsecond resolution necessary to image the disk within a few \unit{\au} of the central star for even the closest YSO targets. Long baseline optical interferometry (LBOI) remains the only feasible technique capable of spatially probing this region. To date, near-infrared interferometric studies have largely employed traditional parametric modeling techniques to broadly survey protoplanetary inner-disk geometry \citep[cf.][]{Monnier+2006,Lazareff+2017,Perraut+2021} and demonstrate time-variability in the inner-disk geometry of individual YSO systems \citep[cf.][]{Kobus+2020,Ganci+2024}. Whereas model-independent image reconstruction has been explored \citep[notably][]{Kluska+2020}, LBOI as a technique has historically been unable to achieve sufficient sensitivity or Fourier coverage needed to properly identify individual disk features. Recent technological improvements at the CHARA array have finally begun to make this possible \citep[especially when augmented by contemporaneous VLTI observations, cf.][]{Ibrahim+2023}.

Herbig Ae star HD~163296 (MWC 275) presents itself as a promising testbed for models of planetary formation. This \qty{1.96(5)}{\solarmass} \citep{Teague+2021, Izquierdo+2022} star hosts one of the nearest-by protoplanetary disk systems, located a mere \qty{100.5 \pm 4}{\parsec} \citep{Brown+2021, Bailer-Jones+2021} from Earth, and is observable from the southern and much of the northern hemisphere. The HD~163296 system is likely mature along its pre-main-sequence lifetime, with an estimated age of \qtyrange{7}{10}{\mega\year} \citep{Setterholm+2018, Guzman-Diaz+2021}; evidence for several planet candidates in its outer disk (beyond \qty{10}{\astronomicalunit}) have been previously claimed in the literature \citep{Pinte+2018,Izquierdo+2022}.

Among published works, no two epochs of high-contrast scattered light coronagraphic imaging of HD~163296's outer disk depict the same shadowing contours \citep{Monnier+2017,Guidi+2018,Mesa+2019,Rich+2020}, even on timescales of under 3 months \citep{Rich+2019}, suggesting considerable activity at closer angular separations. Previous interferometric studies have demonstrated that the orientation of HD~163296's inner disk \citep{Tannirkulam+2008,Benisty+2010,Lazareff+2017,Setterholm+2018,Sanchez-Bermudez+2021} is well co-aligned with the outer disk \citep{Isella+2018,Guidi+2022}, indicating that this source's shadowing is not due to precession of a mis-aligned inner-disk. Infrared interferometry has also directly demonstrated temporal variation within the HD~163296 inner disk \citep{Kobus+2020,Varga+2021,Sanchez-Bermudez+2021}, though studies published to date have lacked the spatial resolution necessary to fully disentangle the relative star/disk flux contributions nor isolate the responsible phenomena and have lacked the temporal resolution necessary to solve for relevant timescales. Without such spatial and temporal resolution, the physical mechanisms responsible for these dramatic changes cannot be adequately determined.

\begin{table*}[p]
    \begin{tabular}{ l c l l c c l }
        \hline\hline
        UT Date & Epoch & Instrument & Configuration & n.\,$\mathcal{V}^2$ & n.\,C.P. & Calibrator(s) \\
        \hline
        2019-06-05   & A &   MIRC-X    &   S1-S2-E1-W1-W2 \hfill (5T)      &   20   &   20   &   HD 168116   \\ % (G8III, H5.430)
        2019-06-06   & B &   MIRC-X    &   S2-E1-E2-W1-W2 \hfill (5T)      &   43   &   50   &   HD 165833, % (G0Ib, H4.966)
        HD 172905   \\
        2019-06-08   & C &   MIRC-X    &   S1-S2-E1-E2-W1-W2 \hfill (6T)   &   30   &   30   &   HD 152418, HD 172905 \\
        2019-06-09   & D &   PIONIER   &   D0-G2-J3-K0 \hfill (Medium)     &   18   &   12   &   HD 162255, % (G3V, H5.683)
        HD 163135   \\ % (K1III, H5.628)   \\ % 0101.C-0896(D)
        2019-06-09   & D &   MIRC-X    &   S1-S2-E1-E2-W1-W2 \hfill (6T)   &   63   &   56   &   HD 172905, % (K0III, H4.948)
        HD 175526   \\ % (K0III, H4.954)
        2019-06-10   & E &   MIRC-X    &   S1-S2-E2-W2 \hfill (4T)   &   12   &    6   &   HD 172905 \\
        2019-07-10   & F &   PIONIER   &   D0-G2-J3-K0 \hfill (Medium)     &   18   &   12   &   HD 154436, % (K0III, H5.265)
        HD 166309   \\ % (G8III, H5.312)   \\ % 0103.C-0915(B)
        2019-07-13   & G &   MIRC-X    &   S1-S2-E2-W1-W2 \hfill (5T)      &   15   &   17   &   HD 156365   \\
        2019-07-20   & H &   PIONIER   &   A0-G1-J2-J3 \hfill (Large)      &   24   &   16   &   HD 154436, HD 166309   \\ % 0103.C-0915(A)
        % \multirow{2}*{2019-07-20}   &  \multirow{2}*{H} &   \multirow{2}*{\mircx}    &   \multirow{2}*{S1-S2-E1-E2-W1-W2\hspace{4mm}\hfill (6T)}   &   \multirow{2}*{92}   &   \multirow{2}*{100}   &   HD 166290, % (K0III, H5.852)
        % HD 156365, \\ % (G3V, H5.197)
        % & & & & & & HD 161037   \\ % (K0III, H6.519)
        2019-07-20   &  H &   \mircx    &   S1-S2-E1-E2-W1-W2\hspace{4mm}\hfill (6T)   &   92   &   100   &   HD 166290, % (K0III, H5.852)
        HD 156365, % (G3V, H5.197)
        HD 161037   \\ % (K0III, H6.519)
        2019-07-29   & all &  PIONIER   &   A0-B2-C1-D0 \hfill (Small)      &   6    &   4    &   HD 154436, HD 166309   \\ % 
        \hline
    \end{tabular}
    \vspace{6pt}
    \caption{Summary of all observations conducted in this campaign. Datasets with the same epoch designation were analyzed together. The \mbox{n.\,$\mathcal{V}^2$} and \mbox{n.\,C.P.} columns refer to the number of statistically independent, averaged square visibility and closure phase measurements, respectively. Each such measurement contains data spanning several wavelength channels.}
    \label{tab:observations}
\end{table*}

\begin{figure*}[p]
    \centering
    \includegraphics[width=0.9\linewidth]{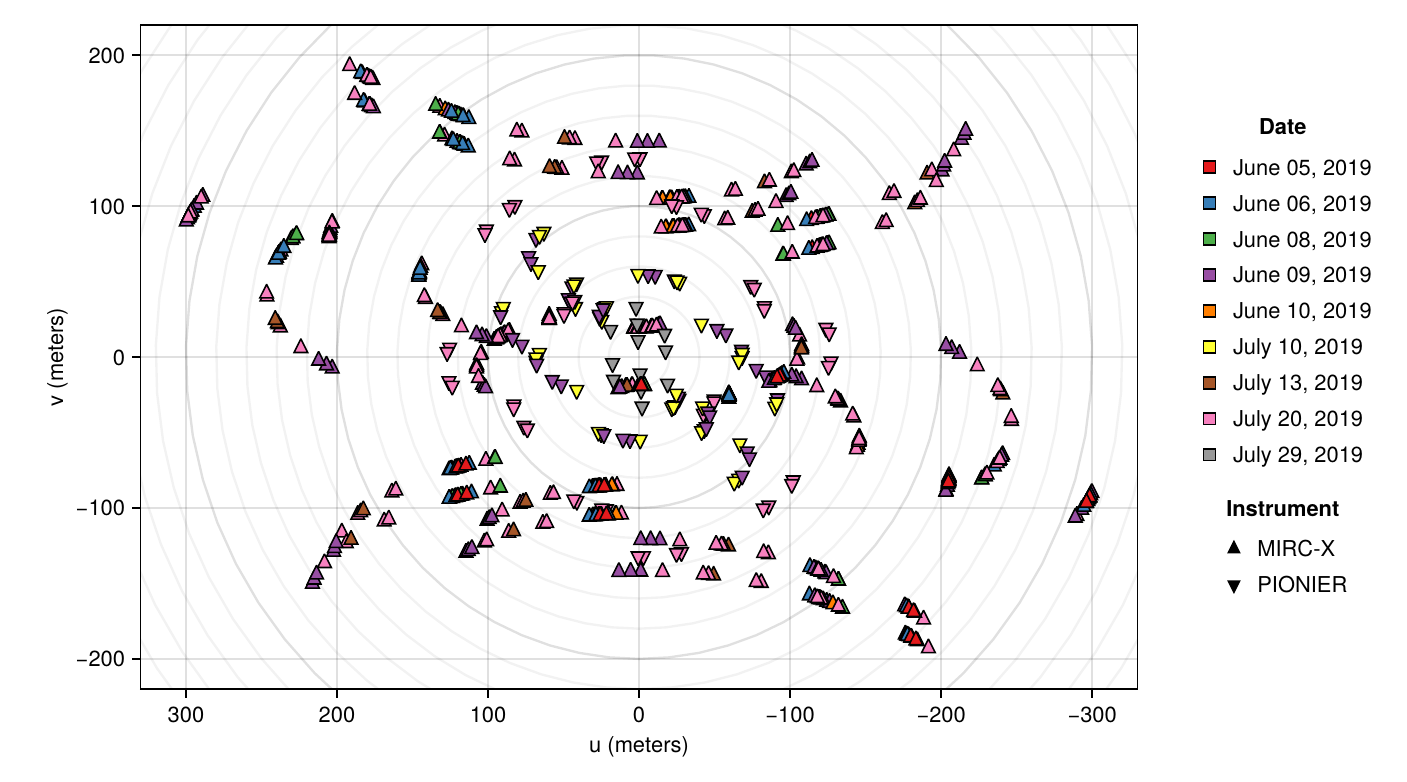}
    \caption{Total $uv$-coverage of the combined \mircx and PIONIER dataset over all examined nights, shown as a function of the projected baseline separation in meters. Note that the CHARA north-south projected baselines are especially foreshortened compared to the east-west baselines, owing to HD~163296's relatively low transit elevation of \ang{34} at CHARA's latitude.}
    \label{fig:uvcoverage}
\end{figure*}

\begin{figure*}[p]
    \centering
    \includegraphics[width=\linewidth]{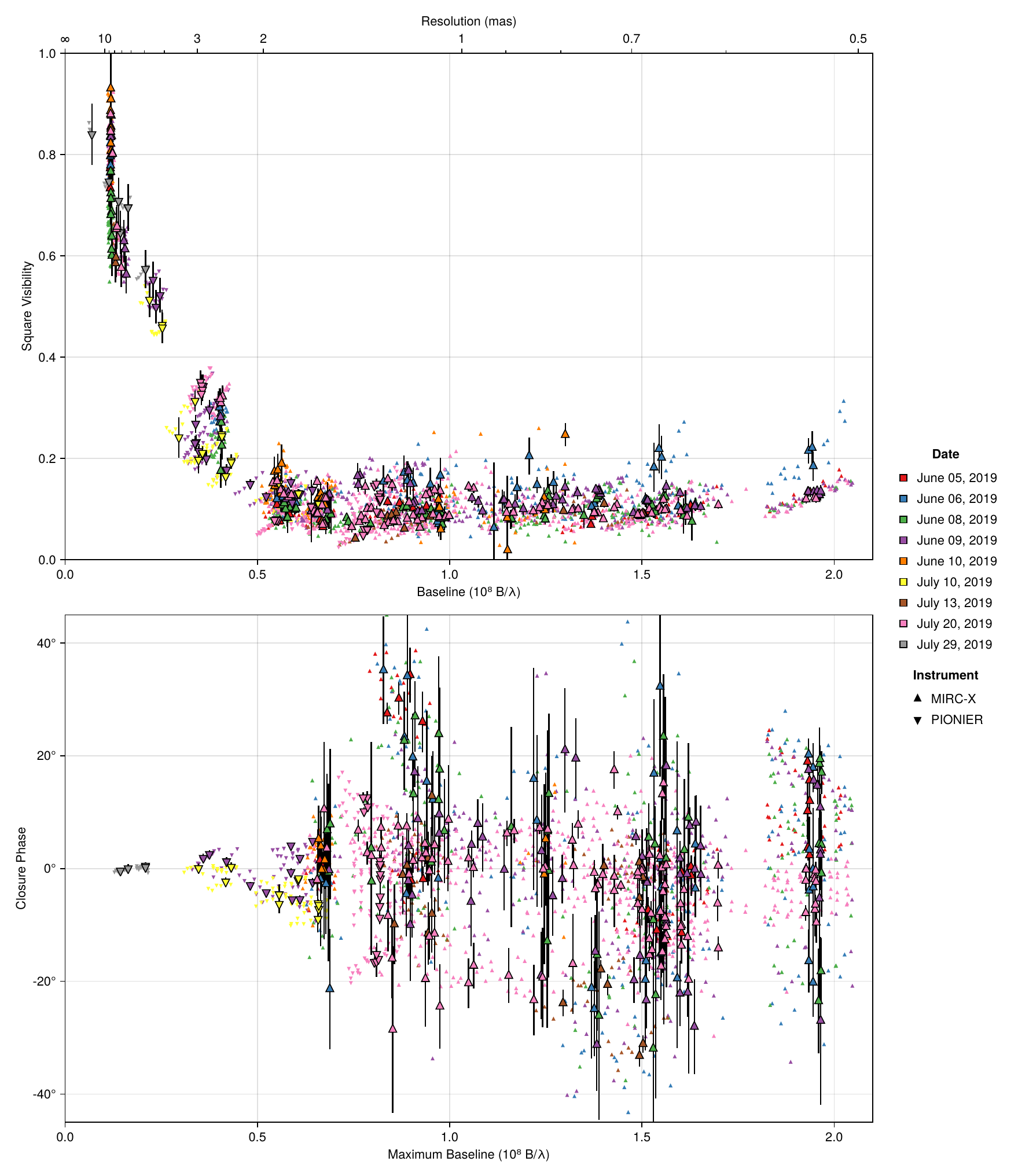}
    \caption{Square visibility (top) and closure phase (bottom) of all analyzed data. The large triangle shapes show the data in the spectral channel nearest \qty{1.6}{\micro\meter} to highlight the number of observations, with the remaining data drawn at a smaller scale to demonstrate the total data volume. Displayed error bars are typical for other wavelengths. The resolution scale at the top indicates the $\theta = \lambda/2B$ scale of the [maximum] baseline separation.}
    \label{fig:rawdata}
\end{figure*}

\section{Observations}

Near-infrared square visibility and closure phase measurements of HD~163296 were collected with the \mircx instrument \citep{Anugu+2020} at the CHARA Array \citep[Mt. Wilson, California;][]{tenBrummelaar+2005} and the PIONIER \citep{LeBouquin+2011} instrument at VLTI (Cerro Paranal, Chile) from early June to mid-July, 2019.\footnote{\mircx observations were collected as parts of the 2019A semester M4 and M11 observing programs; raw data are available to download from the \dataset[CHARA archive]{http://chara.gsu.edu/observers/database}. Program identifiers for the PIONIER dataset, in chronological order, are 0101.C-0896(D), 0103.C-0915(B), 0103.C-0915(A), and 0103.C-0915(C); raw data are available for download from the \dataset[ESO archive]{http://archive.eso.org}.} Both instruments were operated in the \hband (\qtyrange[range-phrase = --]{1.53}{1.73}{\micro\meter}) at a spectral resolution of \mbox{$R \approx 50$}. Altogether, seven nights of observations were collected with \mircx (maximum spatial resolution\footnote{By definition, the maximum spatial resolution $\vartheta$ of an interferometer is determined by the length of its longest baseline $B_\text{max}$ by the relation $\vartheta = \lambda/2B_\text{max}$ (in radians). Given a distance to HD~163296 of almost exactly \qty{100}{\parsec}, physical separations orthogonal to the line of sight are conveniently related to angular sizes with the conversion $\qty{1}{\milliarcsecond} \approx \qty{0.1}{\au}$.} ${\sim}\,\qty{0.5}{\milliarcsecond}$) and three nights with PIONIER where VLTI was set up in the medium (${\sim}\,\qty{1.6}{\milliarcsecond}$) or large (${\sim}\,\qty{1.2}{\milliarcsecond}$) auxiliary telescope configuration. Two of the three PIONIER nights coincided with contemporaneous \mircx observations; for these epochs, the data from both facilities were analyzed together. An additional night of PIONIER measurements with the VLTI array in the small configuration (${\sim}\,\qty{5.0}{\milliarcsecond}$) was conducted at the end of July 2019; these data were included with all previous epochs in order to constrain the flux ratio between the star + inner disk and the surrounding environment \citep{Monnier+2006}. A synopsis of the observations collected in this work is presented in Table~\ref{tab:observations}. The total $uv$-coverage sampled is displayed in Figure~\ref{fig:uvcoverage} and a summary of all (calibrated) interferometric observables used is shown in Figure~\ref{fig:rawdata}.

CHARA data collected on June 8, 2019 and June 10, 2019 were conducted with a polarizing beam-splitter inserted at the end of the \mircx optical path as part of an early experimental run to assess the feasibility of interfero-polarimetric measurements with the instrument \citep{Setterholm+2020}. For the purpose of the present analysis, simultaneous observations of the two measured orthogonal polarization states were averaged together and treated as non-polarimetric data.

All \mircx and PIONIER observations of HD~163296 were alternated with observations of nearby, unresolved calibrator stars having similar brightness ($\left|\Delta H\right| < 1$) to estimate the instrument transfer function. Calibrators were selected with the aid of the JMMC \texttt{SearchCal} service. A summary of these targets' properties are reproduced in Table~\ref{tab:calibrators} for the reader's convenience, with sizes adopted from computed stellar limb darkened diameters in the JMMC Stellar Diameter Catalog, version~2 \citep{Chelli+2016}.

\subsection{Data Reduction}

CHARA fringe measurements were extracted with the \mircx semi-automatic data reduction pipeline \citep{Anugu+2020}, version 1.3.5,\footnote{The \mircx [Python] pipeline is publicly available on the \href{https://gitlab.chara.gsu.edu/lebouquj/mircx_pipeline/}{CHARA GitLab} instance.} and then calibrated and averaged over 10 minute integrations using a custom script adapted from the former MIRC instrument's pipeline.\footnote{The \mircx [IDL] calibration code used in the present analysis is available upon request from \href{mailto:monnier@umich.edu}{John~D.~Monnier}; these routines are currently undergoing translation into Python and will be incorporated into a future version of the official \mircx pipeline.} These routines provided better identification and removal of blocks of data with flux drops, temporary fringe loss, or otherwise poor signal/noise ratio, compared to the automatic calibration program distributed with the current version of the official \mircx pipeline. PIONIER data were similarly reduced, calibrated, and averaged over 15 minute blocks using the instrument's standard data reduction program \texttt{pndrs}.\footnote{The PIONIER [Yorik] pipeline is available from \href{https://www.jmmc.fr/pionier/}{JMMC}.} Reduced OIFITS files for all epochs from both instruments are publicly available for download at \dataset[OiDB]{http://oidb.jmmc.fr/collection.html?id=cb5b2fd1-e23a-4c6c-9b2f-d1e23aec6c95}.

For observations from both facilities, a minimum error of \qty{6.6}{\percent} was applied to all square visibility measurements and an error floor of \ang{0.1} was imposed on all closure phase measurements; these error minima were enforced to account for atmospheric seeing effects, uncertainty in the instrumental transfer function, and other data processing biases \citep{Monnier+2012}. A small number of poor quality square visibility and closure phase measurements, with errors exceeding 0.2 and \ang{20}, respectively, were removed from further analysis.

\begin{table}[t]
    \centering
    \begin{tabular}{ c c c c c }
        \hline\hline
        Object & SpT & V & H & LDD \\
        &   &  (mag) &  (mag) &  (mas)\\
        \hline
        HD~152418 &  K0III &   7.68   &  4.10 &  \num{0.774(53)} \\
        HD~154436 &  K0III &   8.45   &  5.26 &  \num{0.510(12)} \\
        HD~156365 &    G3V &   6.59   &  5.20 &  \num{0.418(10)} \\
        HD~161037 &  K0III &   9.44   &  6.52 &  \num{0.274(07)} \\
        HD~162255 &    G3V &   7.15   &  5.68 &  \num{0.335(08)} \\
        HD~163135 &  K1III &   8.78   &  5.63 &  \num{0.431(11)} \\
        HD~165833 &   G0Ib &   7.84   &  4.97 &  \num{0.508(12)} \\
        HD~166290 &  K0III &   8.32   &  5.85 &  \num{0.356(08)} \\
        HD~166309 &  G8III &   7.61   &  5.31 &  \num{0.444(10)} \\
        HD~168116 &  G8III &   7.65   &  5.43 &  \num{0.424(10)} \\
        HD~172905 &  K0III &   7.36   &  4.95 &  \num{0.553(13)} \\
        HD~175526 &  K0III &   7.51   &  4.95 &  \num{0.552(14)} \\
        \hline
    \end{tabular}
    \vspace{6pt}
    \caption{Summary of calibrator source properties.}
    \label{tab:calibrators}
\end{table}

\section{Parametric Modeling}

The data were first investigated analytically using a set of parametric models with closed-form representations in Fourier and image space, allowing for the characterization of the general features of the on-sky emission despite scarcity of the Fourier coverage; a mathematical review is provided in Appendix~\ref{app:modeling}. Each model was comprised of three geometrical components, a disk profile (described below), a central point source with an $F_\lambda \propto \lambda^{-4}$ spectrum, and a faint, over-resolved ``halo'' component with uniform flux across the field of view (whose relative flux contribution was set as a free parameter). The flux contribution of the disk component was fixed to a power law as a function of wavelength, with the slope and relative flux contribution left as free parameters. For all model fits, the halo component's spectral slope was fixed to zero (as changing this value had next to no discernible change on the final fit results). Altogether, the visibility for each model at each wavelength channel comprised of the sum of each component times its flux contribution, which was then normalized by the sum of all flux contributions.

Model parameters were determined by maximizing a log-likelihood function

\begin{equation}
    \ln(\mathcal{L}) = -\frac{1}{2}\chi^2
\end{equation}

\noindent where

\begin{multline}
    \chi^2 = \sum_i^{N_{\mathcal{V}^2}}\frac{(\mathcal{V}^2_\text{model,i} - \mathcal{V}^2_\text{data,i})^2}{\sigma_{\mathcal{V}^2}^2} \\
    + \sum_j^{N_\text{CP}}\frac{(\text{CP}_\text{model,j} - \text{CP}_\text{data,j})^2}{\sigma_\text{CP}^2}
\end{multline}

\noindent and $N_{\mathcal{V}^2}$ and $N_\text{CP}$ are the total number of square visibility and closure phase measurements (with data in each spectral channel counted individually), respectively. Optimization of model parameters was accomplished using the nested sampling Monte Carlo algorithm MLFriends \citep{Buchner2016,Buchner2019} via the \texttt{UltraNest} software package \citep{Buchner2021}. Fit parameters were allowed to sample the entire space of physically relevant values, and uniform priors were used for all parameters. Each fit was conducted with the minimum number of live points set to 1000.

In our initial exploration, we attempted to fit all measured data together using a number of common geometric models found in the YSO inner-disk literature, including [combinations of] Gaussian disks \citep{Setterholm+2018}, skewed rings \citep{Lazareff+2017,Sanchez-Bermudez+2021,Varga+2021}, and double sigmoids \citep{Ibrahim+2023}. Across the family of tested models, we broadly recover inner-disk orientation and characteristic size parameters consistent with previously published literature for HD~163296. However, none of these ``simple,'' temporally-static models were able to fit well to the higher-density, higher-sensitivity observations acquired in this study compared to past work: even the best models failed to achieve reduced-$\chi^2$ goodness-of-fit measures better than ${\sim}13$.

We next separated our observations into individual-night epochs (plus the PIONIER small-configuration dataset), fitting a series of Gaussian-convolved thin ring models with increasing sinusoidal azimuthual variation, spanning from 0th to 8th order. Our disk models therefore contained the following free parameters for fitting at each epoch: ring radius, Gaussian kernel width, and amplitude and phase of the various azimuthal modulation components. In order to establish a comparative consistency among the various epochs, we fixed the inclination of our model inner disk to \ang{46} and the position angle to \ang{133}, matching the orientation\footnote{Per convention, inclination angle is measured with respect to a face-on orientation and the position angle of the major axis is measured east of north.} parameters measured by ALMA \citep{Guidi+2022}. We note that without this assumed orientation, the fits take significantly longer to converge and that in resulting model fits the azimuthal terms dominate over the orientation terms, producing flux maxima in roughly the same places as in the model fits shown here, but with (unphysical) orientations randomly oriented between epochs and no substantial improvement in final $\chi^2$. Considering that there is no compelling evidence in the literature to suggest that the inner and outer disks in this system are misaligned (see Introduction), fixing the orientation to the ALMA measured parameters is justified.

\begin{figure*}[]
    \centering
    \includegraphics[width=\linewidth]{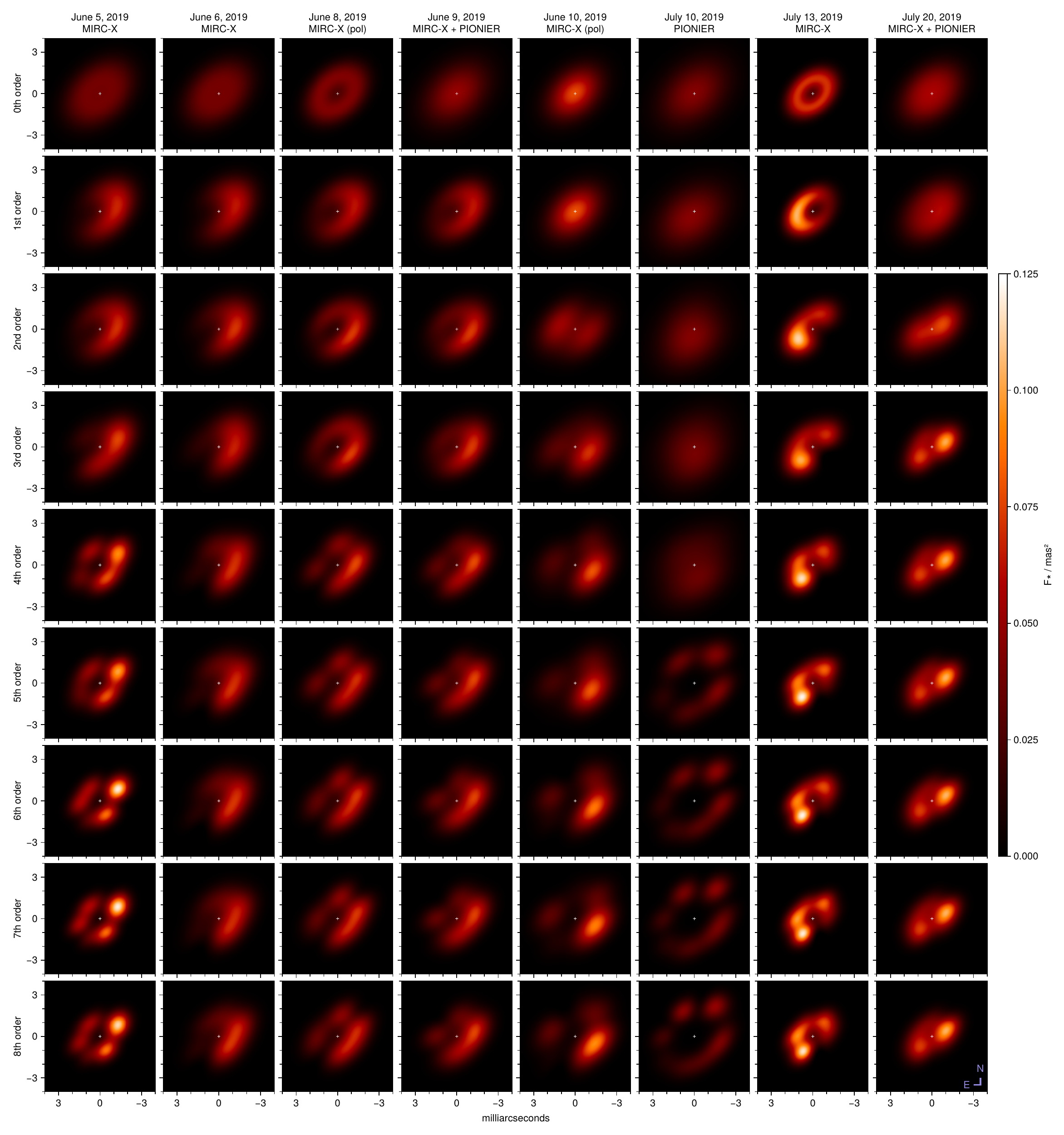}
    \caption{Best fit azimuthally modulated rings for each epoch at \qty{1.6}{\micro\meter}. The orientation of the inner disk was fixed to match the outer disk orientation, but the ring size, and amplitude and phase of sinusoidal variations were free parameters. Each column represents a single epoch of data; each row represents the maximum sinusoidal order used in the fit. For all epochs, models beyond 5th order do not show significant qualitative deviation with the addition of subsequent higher-order sinusoids.}
    \label{fig:gallery}
\end{figure*}

\begin{figure*}[]
    \centering
    \includegraphics[width=\linewidth]{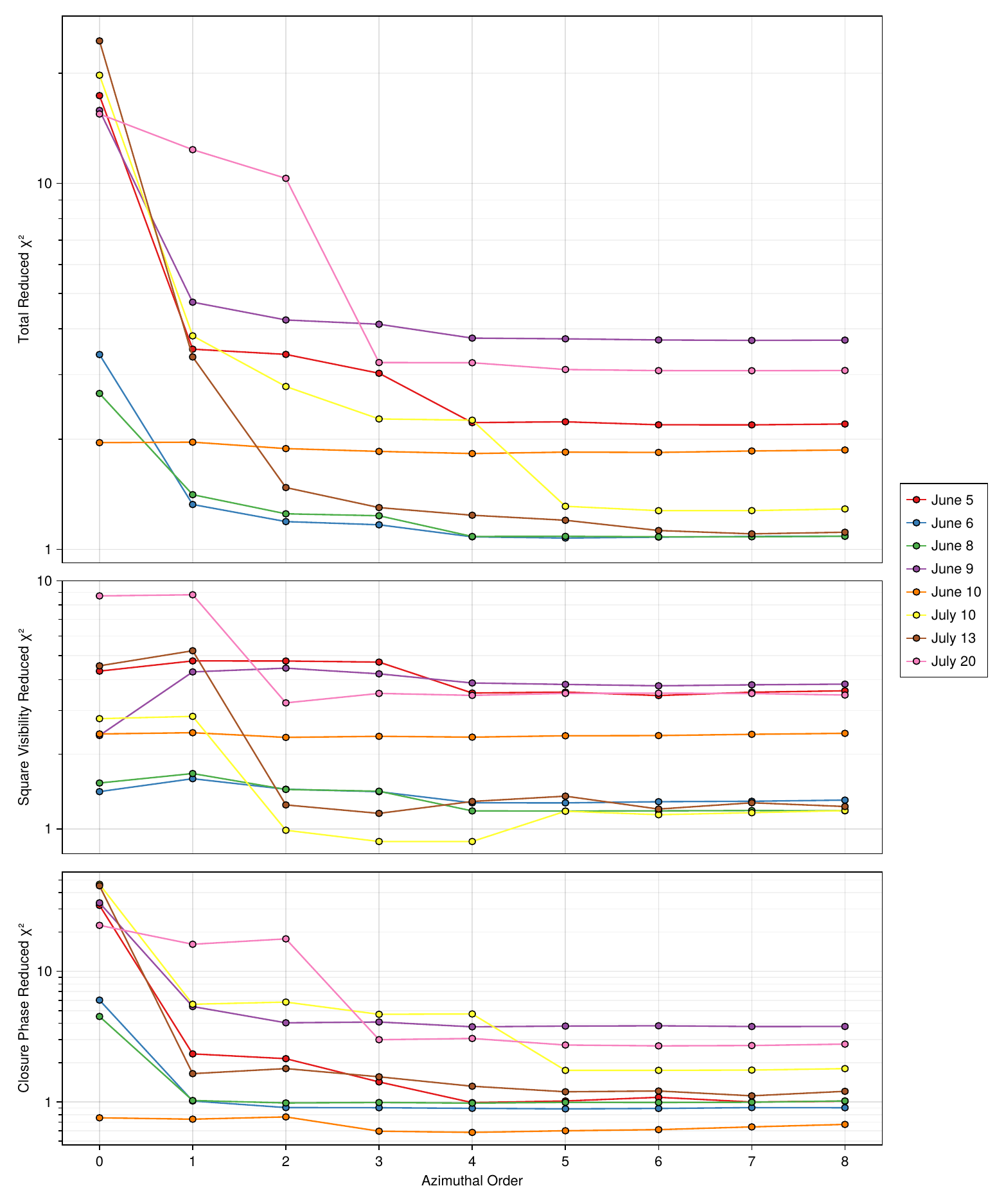}
    \caption{The total $\chi_\nu^2$ value (top panel) and $\chi_\nu^2$ contributions of the square visibility (middle) and closure phase (bottom) components of the analytic modeling fits. The addition of additional azimuthal orders beyond 3--5 has little effect on the final fit for all epochs.}
    \label{fig:order_chi2}
\end{figure*}

A gallery of all data epochs and azimuthal orders is presented in Figure~\ref{fig:gallery}. Across all epochs, fit results generally continued to improve with increasing azimuthal variation up to 5th order, after which the inclusion of additional sinusoidal components no longer optimized the Bayesian information criterion \citep{Schwarz1978} (see also model reduced $\chi^2$ values in Figure~\ref{fig:order_chi2}). We do note however that the July 10th dataset has substantially shorter maximum baselines than all other epochs\footnote{Indeed Figure~\ref{fig:summary_july10} clearly shows that we only sample along the primary visibility lobe, so we cannot in principle confidently extract much more information beyond the representative size scale of the system.}, and the resulting fits above 4th order are clear examples of over-fitting. We largely ignore this data epoch in subsequent discussion owing to its comparatively inferior spatial resolution. Among the other epochs, a typical ring radius of \qty{0.16(3)}{\au} was recovered across the suite of fits.\footnote{Fit parameters for the 5th order azimuthal variation models are provided in Table~\ref{tab:fitparm}.}

Significant variation in azimuthal geometry is retrieved between epochs: most exhibit a dominant ``blob'' along the ring albeit with changing location. The evolution of this dominant blob, especially among the first five epochs which were collected within a span of days, is discussed in detail in \S\ref{sec:discussion}.

\section{Image Reconstruction}

\begin{figure*}[!t]
    \centering
    \includegraphics[width=0.94\textwidth]{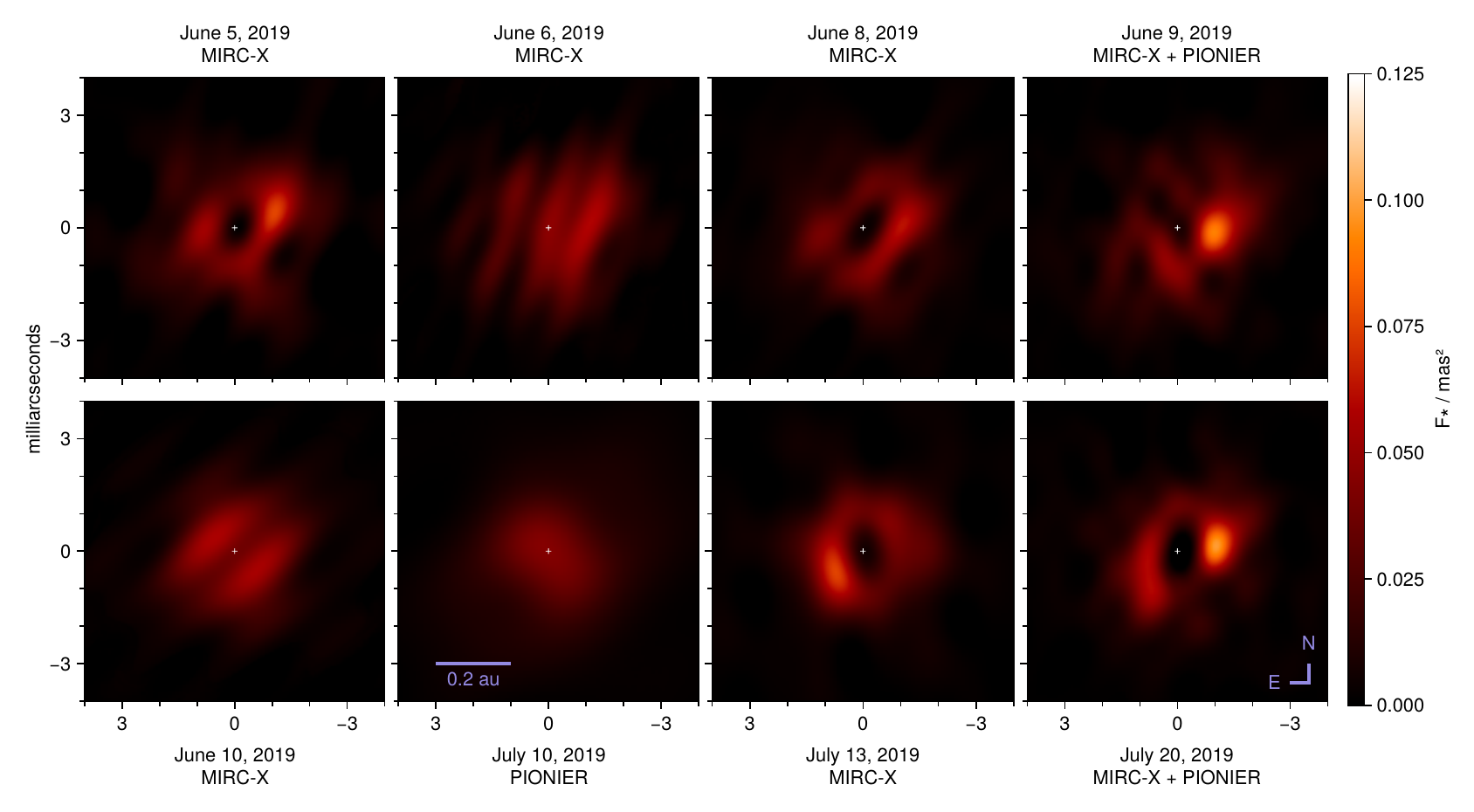}
    \caption{Images reconstructed of the HD~163296 inner disk via \texttt{OITOOLS.jl}. A clear, moving, localized flux excess is recovered near the putative dust sublimation rim. We emphasize that while these images do provide a general sense of the inner geometry, and that the presence and rough position of the localized spot is corroborated by traditional model fitting methods, other ``whispy'' features seen in these images are likely artifacts of the image reconstruction process given the sparse interferometric Fourier coverage rather than real, physical disk features.}
    \label{fig:imaging}
\end{figure*}

In order to test the robustness of this rotating pattern, the data were further investigated using model-independent interferometric image reconstruction.\footnote{We emphasize that, while image reconstruction can broadly constrain the basic geometry of the object being observed and thus provides a better understanding of the overall emission distribution as a whole, these methods are prone to over-fitting data with sparse Fourier coverage often resulting in the presence of non-physical artifacts. Thus, we caution the reader from over interpreting the precise contours in the ``images'' presented later in this section.} Specifically, we used the \texttt{OITOOLS.jl} software package which employs the semi-Newton method VMLMB \citep{Thiebaut2002} to solve the regularized maximum likelihood problem. Gradient-descent results are generally prone to reflect local minima and are heavily dependent on the imaging priors. To assist the algorithm in producing physically reasonable solutions, regularization constraints of compactness (penalty to flux fit near the edge of the reconstructed image) and total square variation (penalty to ``noisy'' pixel patches, but not to sharp boundaries) were imposed. In order to avoid introducing bias into the prior estimation, initial images were generated with randomly generated pixel values; notably, the results of the parametric modeling were \emph{not} used as the starting images in order to evaluate our observations via completely independent methods.

Each image reconstruction was conducted including the additional analytical contributions of a central star (point source) and a uniformly illuminated background ``halo'' component ($\delta$-function in Fourier space). Fit parameters for the fractional flux contributions of these components took into account spectral information in the interferometric measurements utilizing the \texttt{SPARCO} method \citep{Kluska+2014}, whereby a spectral slope of \num{-4} was fixed for the star and the spectral slopes for the image and halo were fit during the image reconstruction procedure. The fitting procedure alternated between the optimization of the image (i.e. the disk) and the relative fluxes of the additional analytical components until the final values converged (typically 3 iterations).

Images were generated across a wide range of pixel scales (\qtyrange{0.01}{0.2}{\milliarcsecond}) and fields of view (\qtyrange{12}{60}{\milliarcsecond}) in order to verify that a consistent picture emerged with each epoch. Regularization hyper-parameters were fine-tuned until the resulting images appeared to strike a balance of suppressing egregious reconstruction artifacts without exhibiting an excess of smoothing or compacting. The same final imaging parameters were applied to all image epochs: the pixel scale was set to 1/16~mas with 256 pixels on each side, compactness was set to a value of $10^9$, and total square variation was set to a value of $10^7$.

Our imaging fits qualitatively recover the same basic disk geometry as the analytical models for most epochs, namely a ring of emission with a similar size and orientation. Moreover, the imaging results consistently recover the presence of a localized, moving disk perturbation with roughly the same position as seen in the analytic model fits (Fig.~\ref{fig:imaging}).

\begin{table*}[!t]
    \centering
    \small
    \begin{tabular}{ l l c c c c }
        \hline\hline
        & & \multicolumn{2}{c}{Analytic Model Fit} & \multicolumn{2}{c}{Image Reconstruction} \\
        Epoch & Instrument(s) & Radius [\unit{\au}] & Phase [\unit{\degree}] & Radius [\unit{\au}] & Phase [\unit{\degree}] \\
        \hline
        2019-06-05   & MIRC-X            & \num{0.15 \pm 0.03} & \num{194 \pm 15} & \num{0.12 \pm 0.02} & \num{208 \pm 19} \\
        2019-06-06   & MIRC-X            & \num{0.13 \pm 0.04} & \num{240 \pm 41} & \num{0.13 \pm 0.03} & \num{234 \pm 32} \\
        2019-06-08   & MIRC-X            & \num{0.15 \pm 0.03} & \num{241 \pm 36} & \num{0.12 \pm 0.02} & \num{230 \pm 22} \\
        2019-06-09   & MIRC-X \& PIONIER & \num{0.14 \pm 0.03} & \num{224 \pm 28} & \num{0.12 \pm 0.02} & \num{238 \pm 14} \\
        2019-06-10   & MIRC-X            & \num{0.17 \pm 0.04} & \num{253 \pm 19} &       -             &      -           \\
        2019-07-10   & PIONIER           &       -             &      -           &       -             &      -           \\
        2019-07-13   & MIRC-X            & \num{0.13 \pm 0.02} & \num{349 \pm 16} & \num{0.09 \pm 0.02} & \num{ 14 \pm 32} \\
        2019-07-20   & MIRC-X \& PIONIER & \num{0.11 \pm 0.03} & \num{209 \pm 21} & \num{0.12 \pm 0.02} & \num{223 \pm 15} \\
        \hline
    \end{tabular}
    \vspace{6pt}
    \caption{Recovered positions of the dominant disk feature, with reported uncertainties based on the \qty{90}{\percent} brightness contours of the best-fit solution. The orbital radius and orbital phase values are extracted assuming that the inner disk is co-oriented with the outer disk as measured by ALMA (inclination $= \ang{46}$, P.A. $= \ang{133}$). Entries are left blank where a unique \qty{90}{\percent} contour solution either does not exist or includes the central star.}
    \label{tab:blob}
\end{table*}

We note that the imaging results for the June 6 and June 10 epochs yield a qualitatively different image compared to the analytic fits; in both cases this appears to be an effect of over-fitting in the image reduction process. For the case of June 10, this is likely the result of data being collected only with 4 operational CHARA telescopes, which significantly reduces the available Fourier coverage. The cause for the June 6 case is a bit more difficult to discern, though we do highlight that the visibilities at long baselines (dominated by the central star's flux contribution) calibrate to substantially higher values compared to all other epochs even after careful, manual re-reduction of the data. We also note that the relatively high extracted uncertainties in the June 6 closure phase measurements (Fig.~\ref{fig:summary_june6}) significantly degrades the determinability of the image reconstruction. Finally, we remind the reader that the July 10 dataset contains only VLTI/PIONIER data and thus has significantly lower resolution and less overall Fourier coverage compared to the other epochs; this data epoch is therefore expected to have poor imaging performance.

\section{Discussion} \label{sec:discussion}

\begin{figure*}[p]
    \centering
    \includegraphics[width=0.94\textwidth]{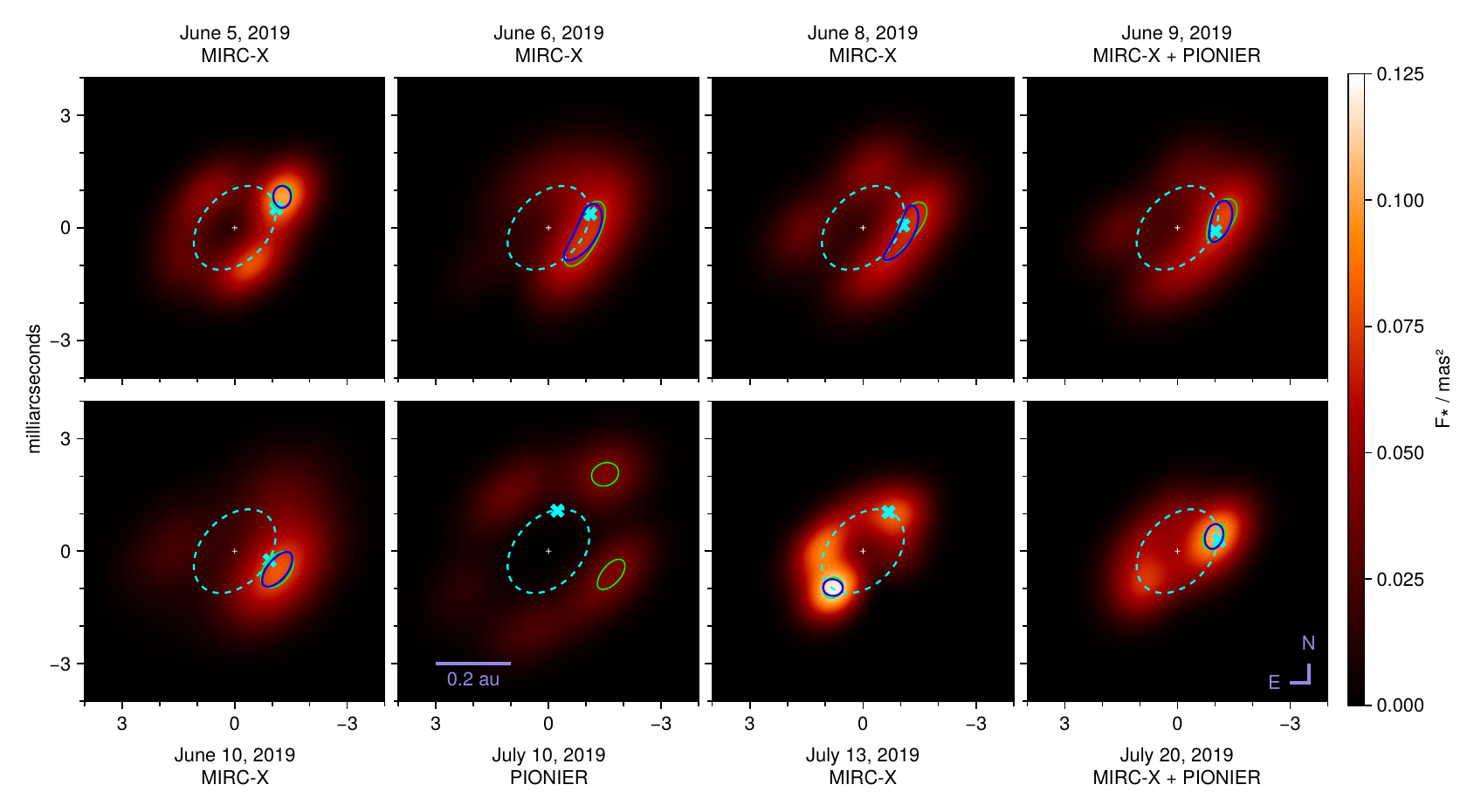}
    \caption{Parametric model (5th order) fits of all analyzed epochs, with the orbital position of a point on a 43.6 day circular trajectory at a radius of \qty{0.13}{\au} superimposed in cyan. The green contours denote the regions in the reconstructed disk image where the surface brightness is within \qty{90}{\percent} of its peak value. The blue lines denote the best-fit error-ellipses to the green contours, which are reported in Table~\ref{tab:blob} and used to fit the cyan orbit. Note that the analytic best-fit model solution for July 13th still places a considerable amount of flux in the vicinity of the expected orbital position, despite the fact that this region is not the dominant source of disk emission in that particular epoch.}
    \label{fig:analytic-orbit}
\end{figure*}

\begin{figure*}[p]
    \centering
    \includegraphics[width=0.94\textwidth]{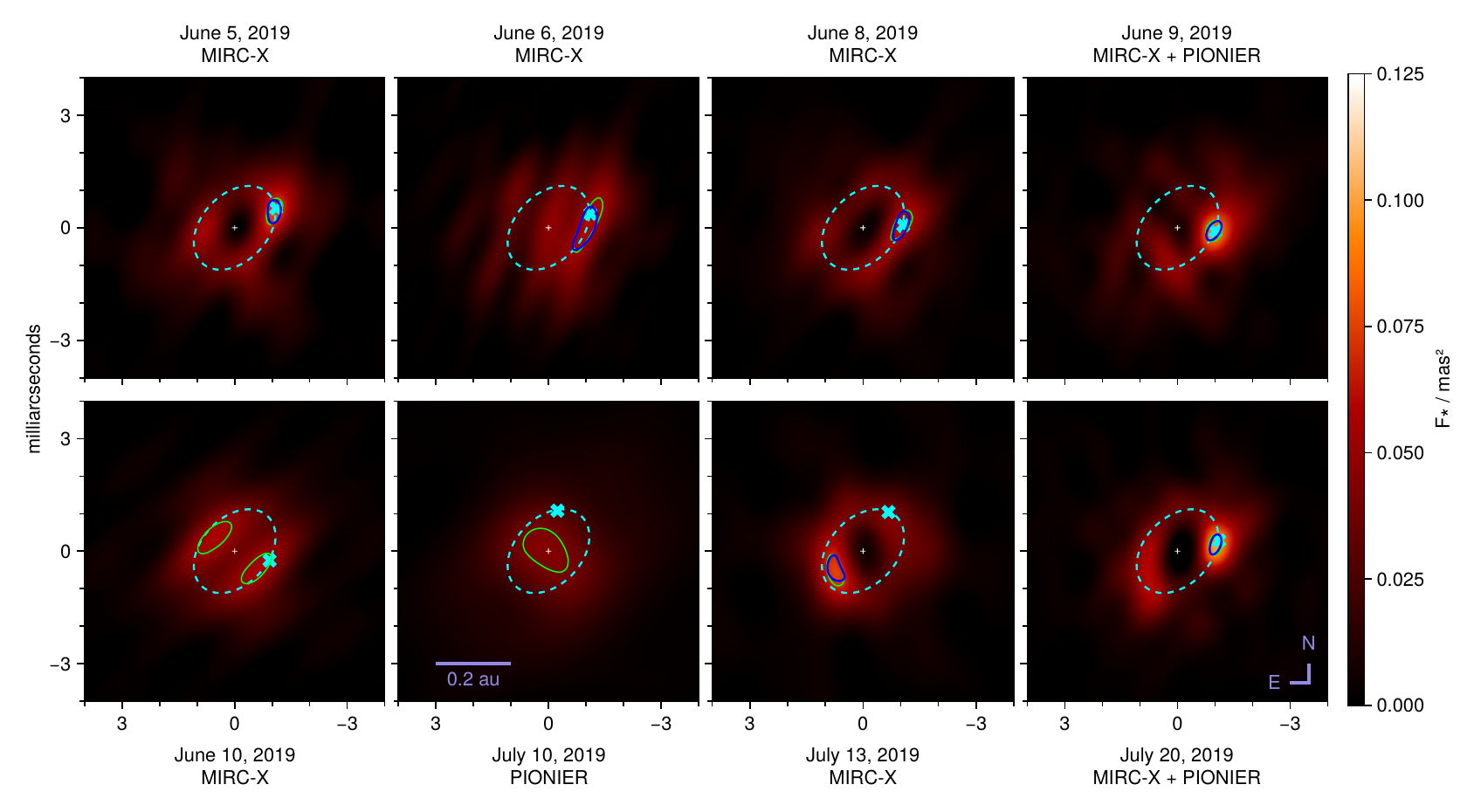}
    \caption{Image reconstructions of all analyzed epochs with orbital parameter extraction diagnostics overplotted (see Fig.~\ref{fig:analytic-orbit} for description).}
    \label{fig:image-orbit}
\end{figure*}

For each epoch in the parametric model fits and image reconstructions, the dominant disk feature position (where possible) is recovered, based on the center of the \qty{90}{\percent} surface brightness contour; this procedure is depicted in Figures \ref{fig:analytic-orbit} and \ref{fig:image-orbit}, respectively. A summary of recovered positions is presented in Table~\ref{tab:blob}. Together, the analytic modeling and the image reconstructions substantiate a moving spot at an orbital radius of \qty{0.13 \pm 0.01}{\au} revolving around the central star on a timescale of \qty{43.6 \pm 3.3}{days}. We note that, while the dominant position recovered on July 13 does not correspond to the extracted orbital location, a secondary localized peak is recovered in the analytical fit and image reconstruction consistent with the aforementioned rotating feature.

While the presence of a dynamic feature within the HD~163296 inner-disk has been recently inferred based on moving shadows observed on the surface of the outer-disk \citep{Rich+2019} as well as with indications of temporal variability seen in VLTI interferometric measurements \citep{Kobus+2020,Varga+2021,Sanchez-Bermudez+2021}, we clearly resolve a discrete spot in locomotion thanks to CHARA's much longer baselines and larger $uv$-coverage footprint. We emphasize the radially localized nature of the recovered ``blob(s),'' with sizes consistent with (and therefore likely exaggerated by) the resolution limit of the interferometric data.

Strikingly, the orbital velocity of the feature is conspicuously sub-Keplerian. This is clearly demonstrated in the early June observations, which contain 5 distinct epochs within 6 consecutive days; in this subset, a clear trend of clockwise motion -- in conjunction with the known preferred orbital direction of the system \citep{Isella+2018} -- is recovered. However, over the course of these 6 days, the disk pattern rotates less than a quarter of a revolution around the central star, manifesting a period roughly three times larger than an expected Keplerian period of \num{12.0 \pm 1.5} days (adopting a central star mass of \qty{2.0 \pm 0.1}{\solarmass}). The next \mircx observations, conducted over a month later, further corroborate this slow extracted period; a distinct disk feature is seen at the expected azimuthal position in both the analytical fit and image reconstructions for all epochs (with the exception of July 10, where only PIONIER measurements are available and whose overall $uv$-coverage is too small to make any definitive statements).

Several phenomena may be possibly responsible for the moving emission pattern observed, though the gravitational influence of a protoplanet is the most likely solution, given the estimated age of the system. A protoplanet located within the inner-disk but beyond the sublimation radius would excite a spiral wake at radial distances of a few scale heights, well beyond the Hill sphere of the planet itself \citep{Bae+2018}. If such a protoplanet is present at an orbital separation of \qty{\sim 0.3}{\au}, the feature we observe could be evidence of such a spiral pattern, increasing the disk scale height at the sublimation radius and moving with the same period as the protoplanet. Furthermore, such a planetary orbit fits well within the expected inner planetary migration limit of \numrange{3}{5} truncation radii, imposed by standing waves in the disk excited by the planetary wake \citep{Miranda+2018}. While we do not directly detect a planet at this radius in our present data set, follow-up observations with \mircx and its new \kband companion beam combiner MYSTIC \citep{Setterholm+2023} -- after forthcoming sensitivity upgrades to these instruments -- may be able to uncover a protoplanetary source if the planet/star contrast ratio is higher than a few percent.

We briefly enumerate several other, non-planetary mechanisms which may be responsible for the observations in descending order of likelihood:

\begin{enumerate}
    \item A stable spiral feature could be formed without the presence of a planet by the precession of gas following elliptical orbits \citep{Bae+2023} near the sublimation front. Such eccentric modes facilitate thermodynamic cooling of the gas and can be sustained for long periods of time, spanning several thousand orbital periods \citep{Li+2021}. In this scenario, a spiral arm would be observed to precess at a slower rate than the local Keplerian velocity \citep{Lee+2019}, and a local overdensity of orbiting gas may explain the apparent change in projected radial separation.
    \item Recent MHD simulations of magnetospheric accretion find that the orbital velocity of gas beyond the magnetospheric truncation radius can be much slower than the local Keplerian velocity \citep{Zhu+2023}. Whereas the feature we resolve is \numrange{10}{20} times further beyond the expected magnetospheric truncation radius for HD~163296, it is still possible we are seeing the effects of sub-Keplerian structures developing as a result of the accretion process such as a one-armed spiral-like feature. The non-uniform dynamic evolution of such features may also explain the seemingly anomalous ``blob'' location in the July 13, 2019 data.
    \item Heating and ionization of silicate grains at the sublimation radius, coupled with the magnetic field of the environment, drives highly turbulent motion at radii immediately beyond the dust evaporation front \citep{Flock+2017}. This results in a local pressure maximum beyond the turbulent region at radii \numrange{\sim 1.5}{2} times the sublimation radius where pebble-sized particles are trapped and protoplanet formation may begin to occur \citep{Flock+2019}. In this dead-zone, the Rossby wave instability produces localized vorticies, which locally increases the vertical magnetic field strength and thus the local scale height of the disk. Due to additional material exposed to and reflecting the incident starlight, it is anticipated that such a vortex would produce an observable local near-infrared disk flux excess of up to \qty{20}{\percent} \citep{Flock+2017}. While such a vorticies do not necessarily move at exactly the local orbital velocity, the less than half-Keplerian pattern speed we observe makes this explanation difficult to justify.
\end{enumerate}

Future follow-up studies of HD~163296 will help discriminate between the possible physical mechanisms responsible for producing the moving pattern. Regardless of whether the spot is due to the presence of a [proto]planet, a coordinated observation campaign of the inner disk with LBOI imaging conducted in conjunction with scattered light coronagraphic imaging of the outer disk shadowing, distributed across several simultaneous nights within the same month, could help to constrain the vertical geometry of the disk feature as well as the inner-rim in general.

There is no reason to presume that the inner-disk of HD~163296 is unusual among the population of nearby protoplanetary systems. (Indeed, asymmetries in at least two other YSO objects have recently been shown to exhibit possible sub-Keplerian motion \citep{Ibrahim+2023,Ganci+2024} within the inner disk, though neither of these studies tracked the locomotion with a cadence of several observations within a complete orbit.) Whereas LBOI studies in the past have typically revisited YSO targets at intervals with many months or years between observations, with the assumption that the measured inner-disk flux distribution remains quasi-static over time, we clearly demonstrate that this is not the case for HD~163296. Now that LBOI as a technique has matured such that a single observation block produces enough interferometric data necessary to conduct image reconstruction, especially when simultaneous measurements from multiple facilities can be used in conjunction with each other, other protoplanetary disk systems will need to be monitored in similar fashion.

\begin{acknowledgments}
The authors would like to personally thank Jean-Baptiste Le Bouquin for his major contributions to the development of the \mircx hardware and software, including the data reduction pipeline.

BRS\ and JDM\ acknowledge support from the NASA FINESST grant (Grant No.\ 80NSSC19K1530).

SK acknowledges support from the European research Council through ERC Starting Grant (Grant Agreement No.\ 639889) and ERC Consolidator Grant (Grant Agreement ID 101003096), as well as through STFC Consolidated Grant (ST/V000721/1).

This work is based upon observations obtained with the Georgia State University Center for High Angular Resolution Astronomy Array at Mount Wilson Observatory.  The CHARA Array is supported by the National Science Foundation under Grant No. AST-1636624 and AST-1715788.  Institutional support has been provided from the GSU College of Arts and Sciences and the GSU Office of the Vice President for Research and Economic Development.

\mircx received funding from the European Research Council (ERC) under the European Union's Horizon 2020 research and innovation programme (Grant No. 639889). JDM acknowledges funding for the development of \mircx (NASA-XRP NNX16AD43G, NSF-AST 1909165).

This research has made use of the \href{https://www.jmmc.fr/pionier/}{PIONIER data reduction package}, \href{https://www.jmmc.fr/aspro/}{\texttt{Aspro2} software}, and the \href{https://www.jmmc.fr/searchcal/}{\texttt{SearchCal} service}, provided by the Jean-Marie Mariotti Center.
\end{acknowledgments}

\vspace{5mm}
\facilities{CHARA (MIRC-X), VLTI (PIONIER)}

\bibliography{citations}{}
\bibliographystyle{aasjournal}

\appendix

\section{Analytical Models}\label{app:modeling}

Geometric models were composed of a series of primitive components, each with analytic or semi-analytic functional forms describing their behavior in Fourier space. Generically, for a rotationally symmetric flux distribution in image space with functional form $\mathcal{F}(r)$, the normalized visibility has the form

\begin{equation}\label{eq:vis}
    \mathcal{V}(u, v) = \mathcal{H}_0 (u,v)
\end{equation}

\noindent where

\begin{equation}\label{eq:normhankel}
    \mathcal{H}_n (u, v) = \frac{\int_0^\infty \mathcal{F}(r)\,J_n(2 \pi r \sqrt{u^2 + v^2})\,r\,dr}{\int_0^\infty \mathcal{F}(r)\,r\,dr}
\end{equation}

\noindent and $J_n$ is the Bessel function of the first kind of order $n$. Next, to introduce azimuthal asymmetries our models, we include several orders of sinusoidal modulation. For a flux distribution with the form

\begin{equation}\label{eqn:azmod}
    \mathfrak{F}(r, \phi) = \mathcal{F}(r) \cdot \sum_{n=0}^k a_n \cos(n\phi + \varphi_n),
\end{equation}

\noindent with $a_0$ equal to unity, the complex visibility takes the form

\begin{equation}\label{eqn:compvis}%{multline}
    \mathcal{V}(u,v) = \sum_{n=0}^k a_n (-j)^n \mathcal{H}_n(u,v) %\\
    \cos\left[n \left(\arg(u, v) + \varphi_n - \frac{\pi}{2}\right)\right].
\end{equation}%{multline}

We approximated projection effects at an inclination $i$ and with the major axis rotated by a position angle $\theta$ (east of north) by transforming the $(u,v)$ coordinates to an effective baseline $(u', v')$ given by

\begin{equation}\label{eqn:twincl}
    \begin{pmatrix}u'\\v'\end{pmatrix} = \begin{pmatrix}
    \cos\theta \cos i & -\sin\theta \cos i \\
    \sin\theta & \cos\theta
    \end{pmatrix}
    \begin{pmatrix}u\\v\end{pmatrix}.
\end{equation}

\noindent Under this transformation, azimuthal modulation components on sky are visible clockwise relative to $\theta$.

For our particular tested model, we began with a simple infinitely-thin ring model:

\begin{equation}\label{eqn:fluxring}
    \mathfrak{F}(r) = \frac{\delta(r - r_0)}{2\pi r_0}
\end{equation}

\noindent with a functional form in visibility space of

\begin{equation}
    \mathcal{V}(u, v) = J_0(2\pi r_0 \sqrt{u^2 + v^2}).
\end{equation}

\noindent Azimuthal modulation was then applied for increasing sinusoidal orders according to Eqn.~\ref{eqn:azmod}, followed by convolution with a Gaussian kernel in image space, which is equivalent to a multiplication by a Gaussian in visibility space. The FWHM of the Gaussians in image and visibility space are related by:

\begin{equation}\label{eqn:gauss}
    \Theta_\mathcal{V} = \frac{4 \ln 2}{\pi \Theta_\mathcal{F}}.
\end{equation}

\noindent The transformation into effective baseline coordinates, according to Eqn.~\ref{eqn:twincl}, was conducted after the Gaussian convolution step.

In addition to the simple Gaussian convolved thin ring models, we also tested a double sigmoid model \citep{Ibrahim+2023} for several sample epochs and azimuthal order combinations, wherein we approximated the integrals in Equation~\ref{eq:normhankel} using trapezoidal numerical integration over radii spanning \qtyrange{0}{20}{\milliarcsecond} at \qty{0.1}{\milliarcsecond} intervals. For each such model attempted, we recovered a radial distribution qualitatively similar to the simple model showing no significant improvement to the final model fit statistics. Since the double-sigmoid radial fits were 2-3 orders of magnitude more computationally expensive to conduct, the analysis in this work was limited to the simpler, Gaussian convolved ring model. We also verified that allowing the orientation parameters to vary did not significantly affect the qualitative fit results.

\begin{table*}
    \centering
    % \begin{rotatetable}
    \begin{tabular}{ r r r r r r r r r r r r r r r r }
        \hline\hline
        \multicolumn{1}{c}{Epoch} & \multicolumn{1}{c}{$r_0$} & \multicolumn{1}{c}{$a_1$} & \multicolumn{1}{c}{$\varphi_1$} & \multicolumn{1}{c}{$a_2$} & \multicolumn{1}{c}{$\varphi_2$} & \multicolumn{1}{c}{$a_3$} & \multicolumn{1}{c}{$\varphi_3$} & \multicolumn{1}{c}{$a_4$} & \multicolumn{1}{c}{$\varphi_4$} & \multicolumn{1}{c}{$a_5$} & \multicolumn{1}{c}{$\varphi_5$} & \multicolumn{1}{c}{$\Theta_\mathcal{F}$} & \multicolumn{1}{c}{$F_\text{disk}$} & \multicolumn{1}{c}{$x_\text{disk}$} & \multicolumn{1}{c}{$F_\text{halo}$} \\
        \hline
        2019-06-05 & \num{1.64} & \num{0.51} & \ang{135.1} & \num{0.22} & \ang{125.7} & \num{0.53} & \ang{50.1} & \num{1.00} & \ang{76.6} & \num{0.32} & \ang{14.6} & \num{1.61} & \num{1.71} & \num{-0.36} & \num{0.30} \\
        2019-06-06 & \num{1.71} & \num{0.93} & \ang{136.8} & \num{0.45} & \ang{105.4} & \num{0.34} & \ang{92.2} & \num{1.00} & \ang{74.4} & \num{1.00} & \ang{71.0} & \num{2.15} & \num{1.54} & \num{-0.46} & \num{0.24} \\
        2019-06-08 & \num{1.79} & \num{0.59} & \ang{135.3} & \num{0.44} & \ang{105.2} & \num{0.27} & \ang{88.7} & \num{0.99} & \ang{51.4} & \num{0.90} & \ang{7.4} & \num{1.92} & \num{1.66} & \num{-0.93} & \num{0.36} \\
        2019-06-09 & \num{1.65} & \num{0.68} & \ang{122.5} & \num{0.34} & \ang{122.4} & \num{0.23} & \ang{50.7} & \num{1.00} & \ang{47.9} & \num{0.94} & \ang{17.0} & \num{1.99} & \num{1.59} & \num{0.63} & \num{0.36} \\
        2019-06-10 & \num{1.93} & \num{0.98} & \ang{119.7} & \num{0.94} & \ang{113.0} & \num{0.90} & \ang{91.4} & \num{0.99} & \ang{16.2} & \num{0.76} & \ang{25.2} & \num{2.42} & \num{1.96} & \num{0.06} & \num{0.19} \\
        2019-07-10 & \num{2.61} & \num{0.38} & \ang{159.4} & \num{0.32} & \ang{70.8} & \num{0.17} & \ang{90.0} & \num{0.52} & \ang{10.5} & \num{1.00} & \ang{45.0} & \num{1.80} & \num{1.03} & \num{-2.06} & \num{0.14} \\
        2019-07-13 & \num{1.36} & \num{0.40} & \ang{316.4} & \num{0.66} & \ang{0.8} & \num{0.47} & \ang{33.0} & \num{0.94} & \ang{23.3} & \num{1.00} & \ang{26.7} & \num{1.50} & \num{1.71} & \num{-0.51} & \num{0.37} \\
        2019-07-20 & \num{1.29} & \num{0.25} & \ang{123.5} & \num{0.74} & \ang{153.5} & \num{0.86} & \ang{19.0} & \num{0.26} & \ang{59.9} & \num{0.99} & \ang{16.3} & \num{1.73} & \num{1.53} & \num{0.44} & \num{0.57} \\
        \hline
    \end{tabular}
    \vspace{6pt}
    \caption{Best-fit parameters for the 5th order analytical models. For all fits, the inclination and position angle of the disk were fixed to \ang{46} and \ang{133}, respectively. The ring radius (in mas) is denoted $r_0$ following Equation~\ref{eqn:fluxring}. Azimuthal parameters $\{a_n, \varphi_n\}$ are denoted as in Equation~\ref{eqn:compvis}. The Gaussian convolution kernel width (in mas) is denoted $\Theta_\mathcal{F}$ as in Equation~\ref{eqn:gauss}. Relative flux contribution parameters $F$ are set at a reference wavelength, $\lambda_0$, of \qty{1.6}{\micro\meter}, with the star's flux at the reference wavelength set to unity and spectral slope set to \num{-4}. The fit spectral slope of the disk is denoted $x_\text{disk}$; the halo's spectral slope was fixed to 0.}
    \label{tab:fitparm}
    % \end{rotatetable}
\end{table*}

\section{Additional Figures}

\begin{figure*}
    \centering
    \includegraphics[width=\textwidth]{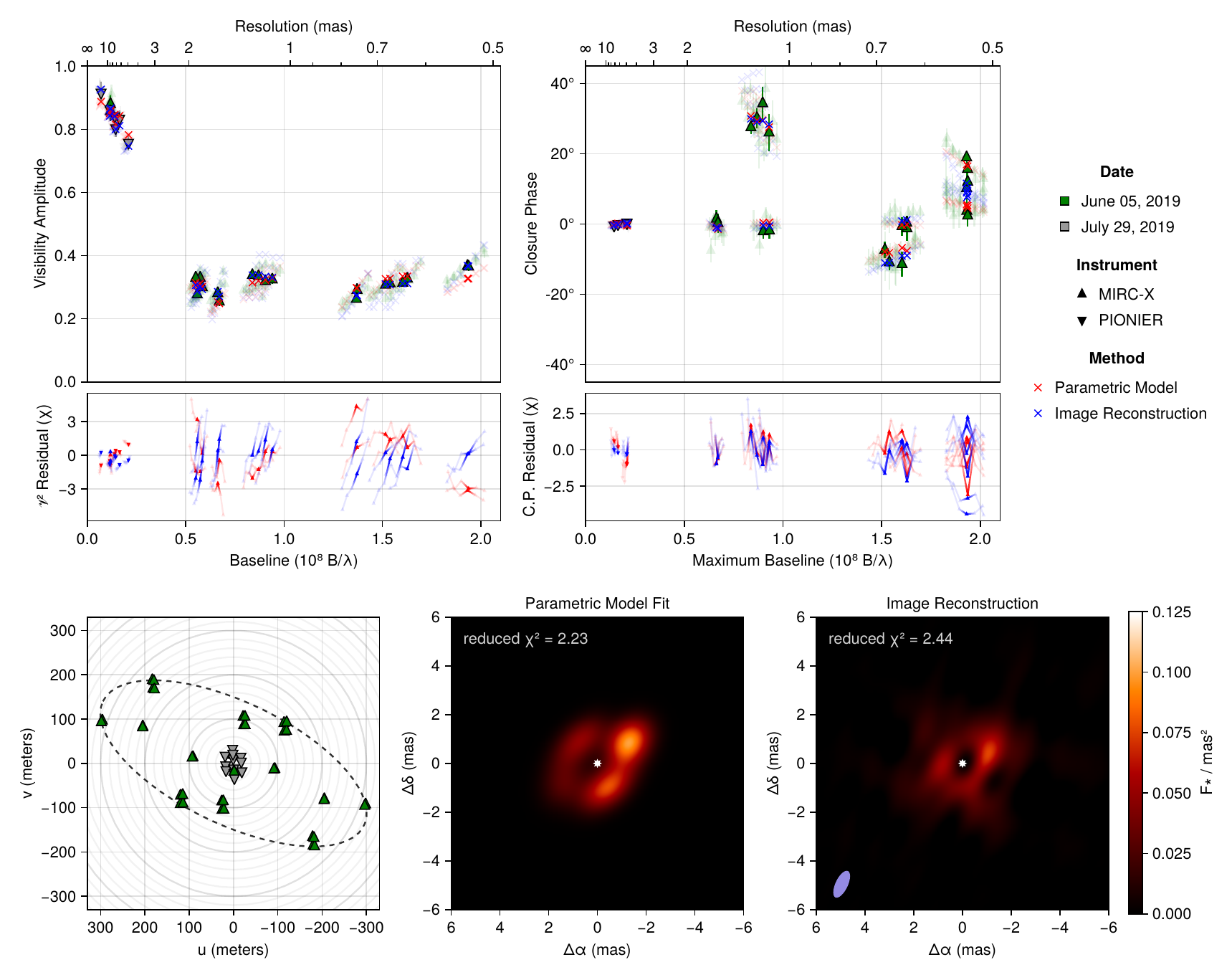}
    \caption{Summary of observations and model fits to the June 5, 2019 (\mircx) epoch data. In the top panels, data points are denoted with triangles with 1-$\sigma$ error bars included. Measurements (and fits) in the waveband closest to \qty{1.6}{\micro\meter} are drawn with full opacity whereas measurements in other wavebands are partially transparent to minimize the ``confusion-limit'' of the figure. In the top left panel, visibility amplitude is shown rather than the measured (and fit to) square visibility to assist the reader in visually distinguishing the halo/disk/star relative flux contributions; the residuals, plotted immediately below, are for square visibility. In the bottom left panel, the total $uv$-coverage of the epoch is shown together with the minimum-area [L\"owner] ellipse enclosing all of the points; this ellipse is used to derive an approximate ``beam-size'' according to $\lambda/2B$ along the major and minor axes, as depicted by the purple ellipse in the bottom left corner of the lower right panel.}
    \label{fig:summary_june5}
\end{figure*}

\begin{figure*}
    \centering
    \includegraphics[width=\textwidth]{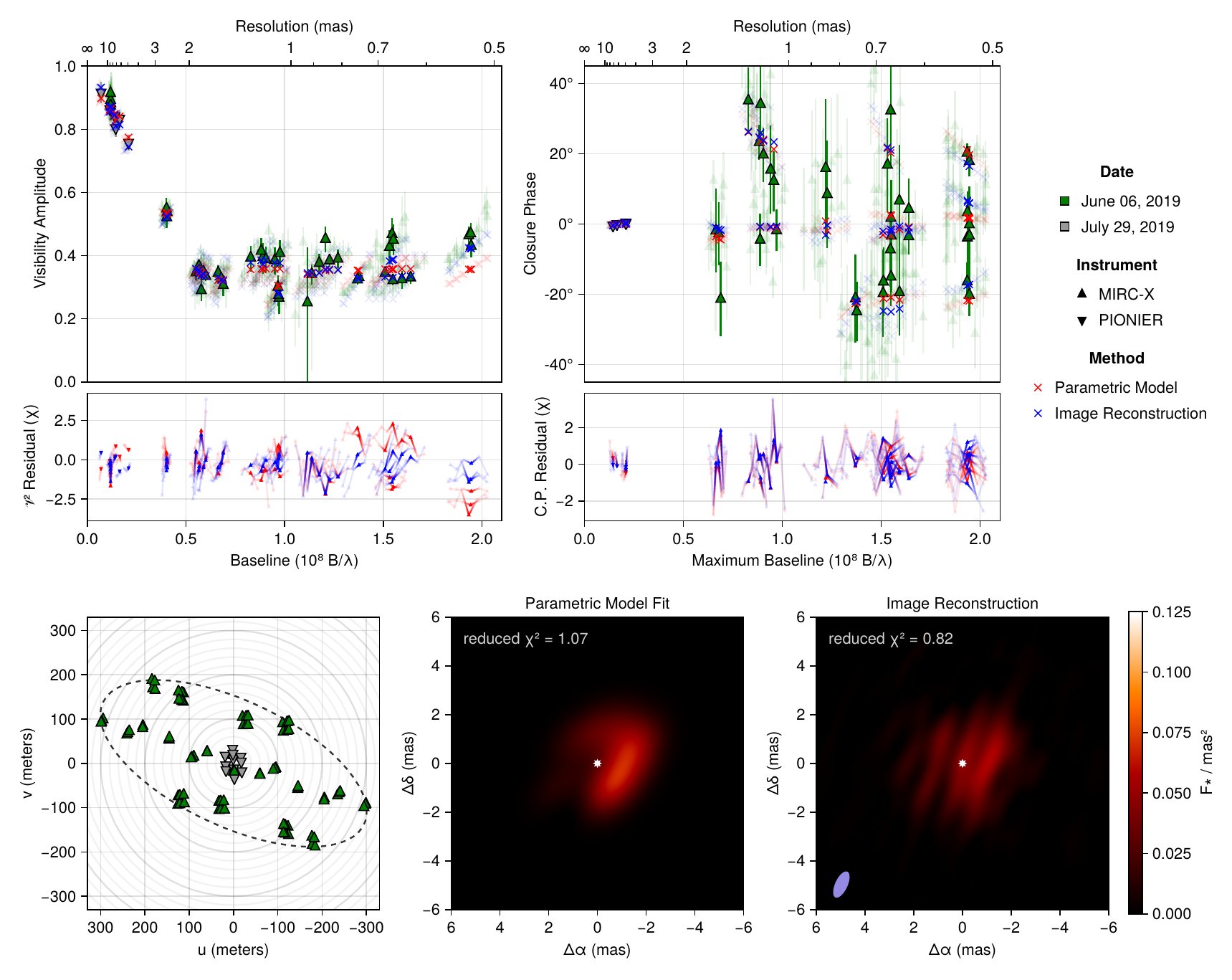}
    \caption{Summary of observations and model fits to the June 6, 2019 (\mircx) epoch data. See the caption in Fig~\ref{fig:summary_june5} for an explanation of what is depicted in the panels.}
    \label{fig:summary_june6}
\end{figure*}

\begin{figure*}
    \centering
    \includegraphics[width=\textwidth]{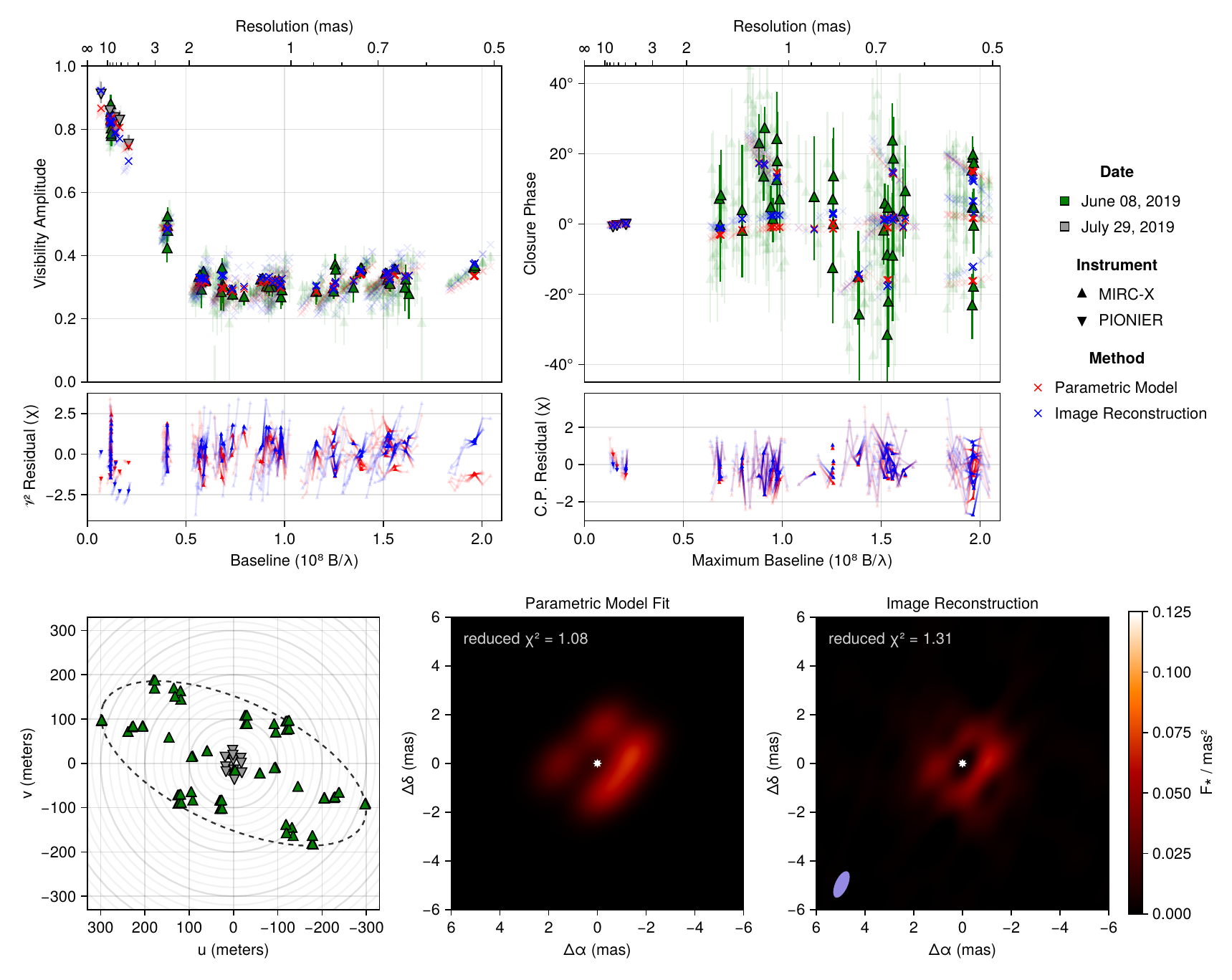}
    \caption{Summary of observations and model fits to the June 8, 2019 (\mircx in polarimetric mode) epoch data. See the caption in Fig~\ref{fig:summary_june5} for an explanation of what is depicted in the panels.}
    \label{fig:summary_june8}
\end{figure*}

\begin{figure*}
    \centering
    \includegraphics[width=\textwidth]{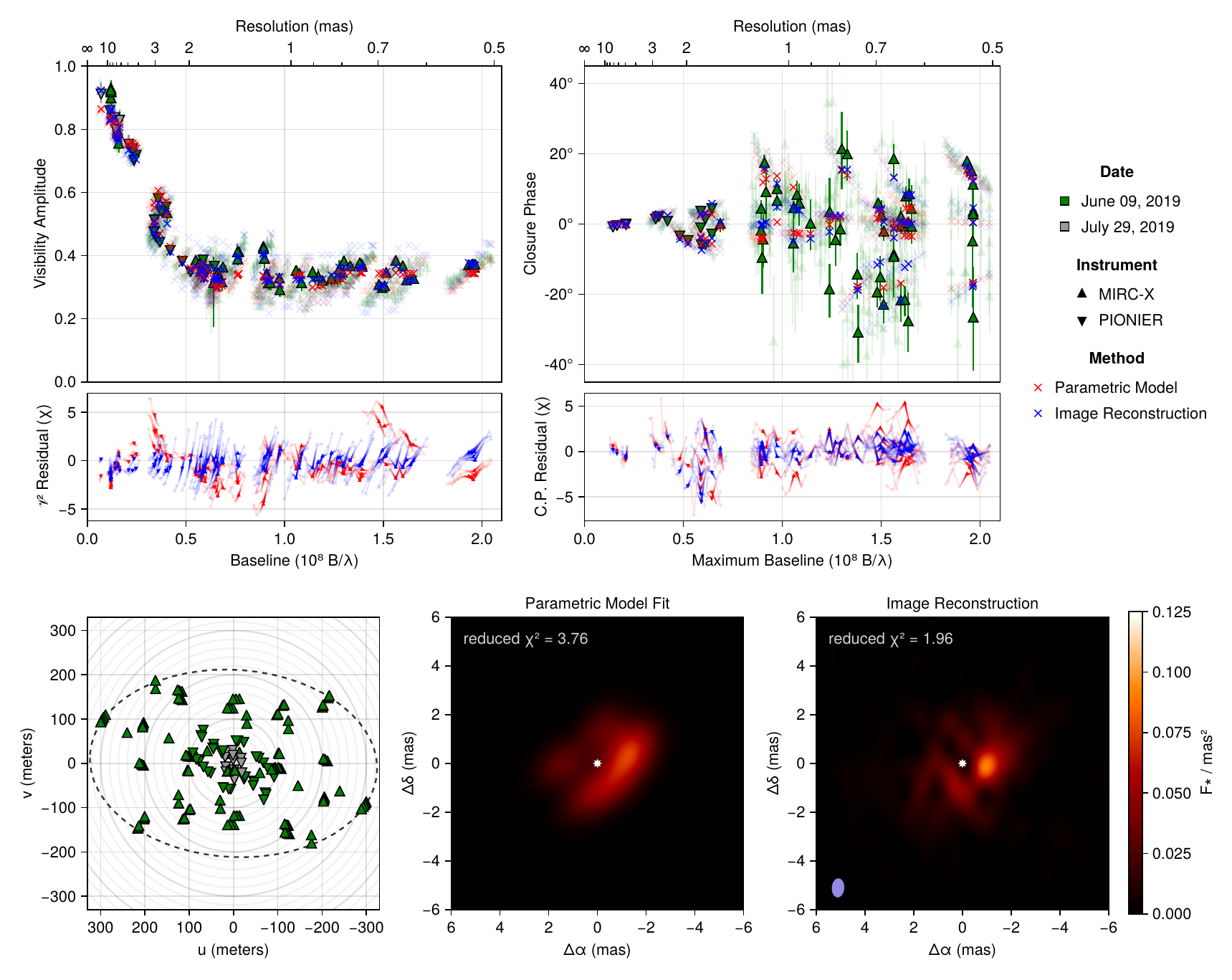}
    \caption{Summary of observations and model fits to the June 9, 2019 (\mircx and PIONIER) epoch data. See the caption in Fig~\ref{fig:summary_june5} for an explanation of what is depicted in the panels.}
    \label{fig:summary_june9}
\end{figure*}

\begin{figure*}
    \centering
    \includegraphics[width=\textwidth]{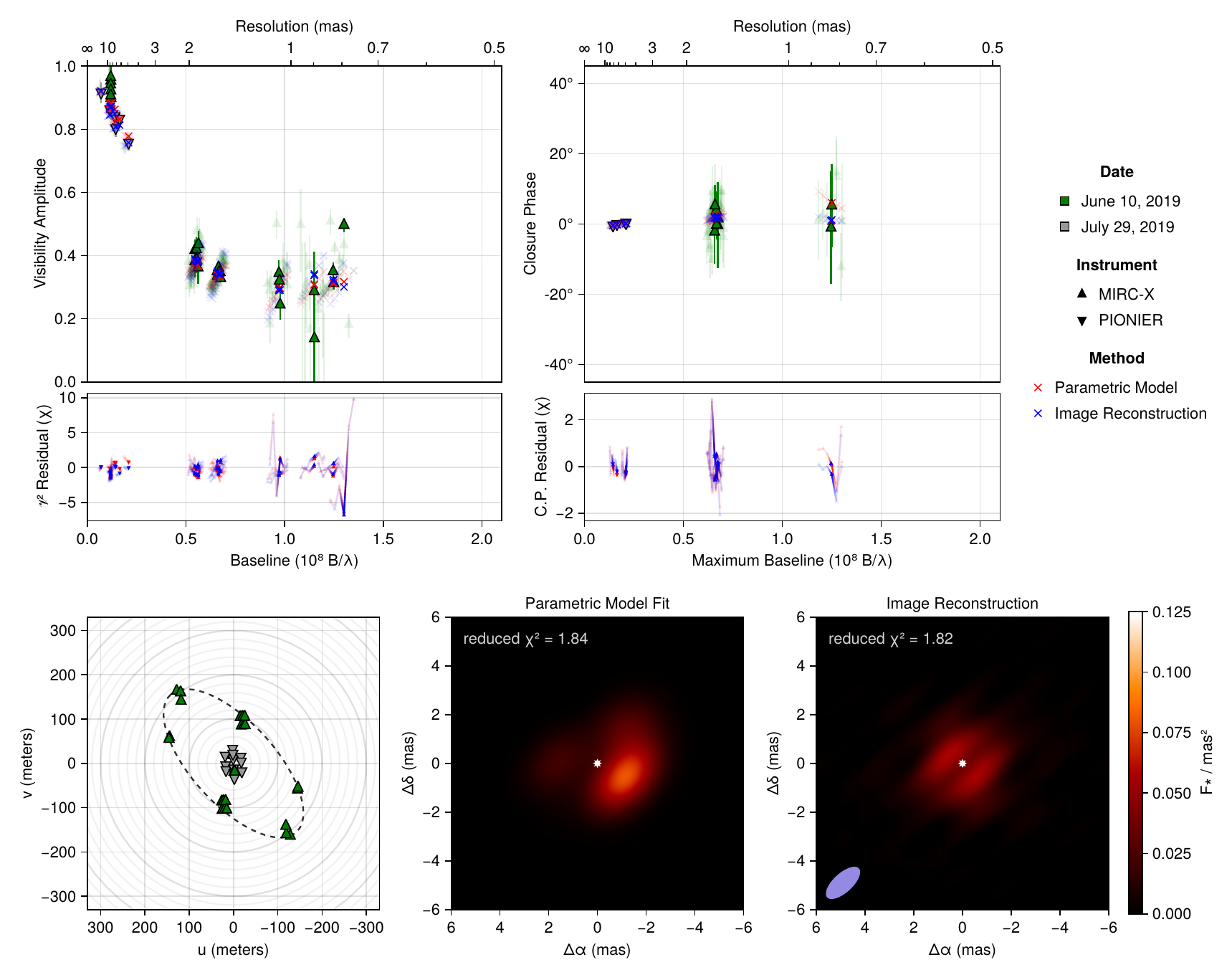}
    \caption{Summary of observations and model fits to the June 10, 2019 (\mircx in polarimetric mode) epoch data. See the caption in Fig~\ref{fig:summary_june5} for an explanation of what is depicted in the panels.}
    \label{fig:summary_june10}
\end{figure*}

\begin{figure*}
    \centering
    \includegraphics[width=\textwidth]{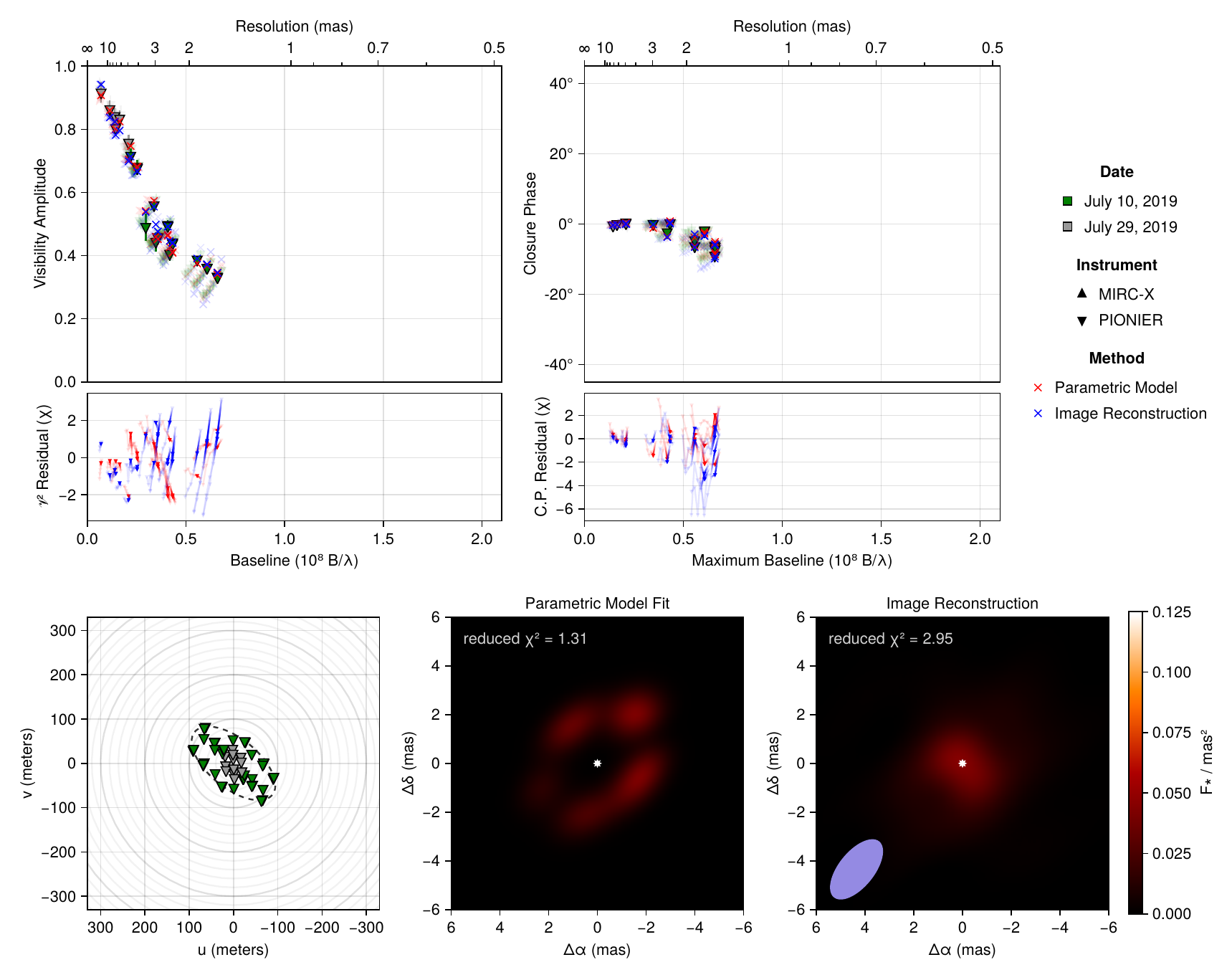}
    \caption{Summary of observations and model fits to the July 10, 2019 (PIONIER only) epoch data. See the caption in Fig~\ref{fig:summary_june5} for an explanation of what is depicted in the panels.}
    \label{fig:summary_july10}
\end{figure*}

\begin{figure*}
    \centering
    \includegraphics[width=\textwidth]{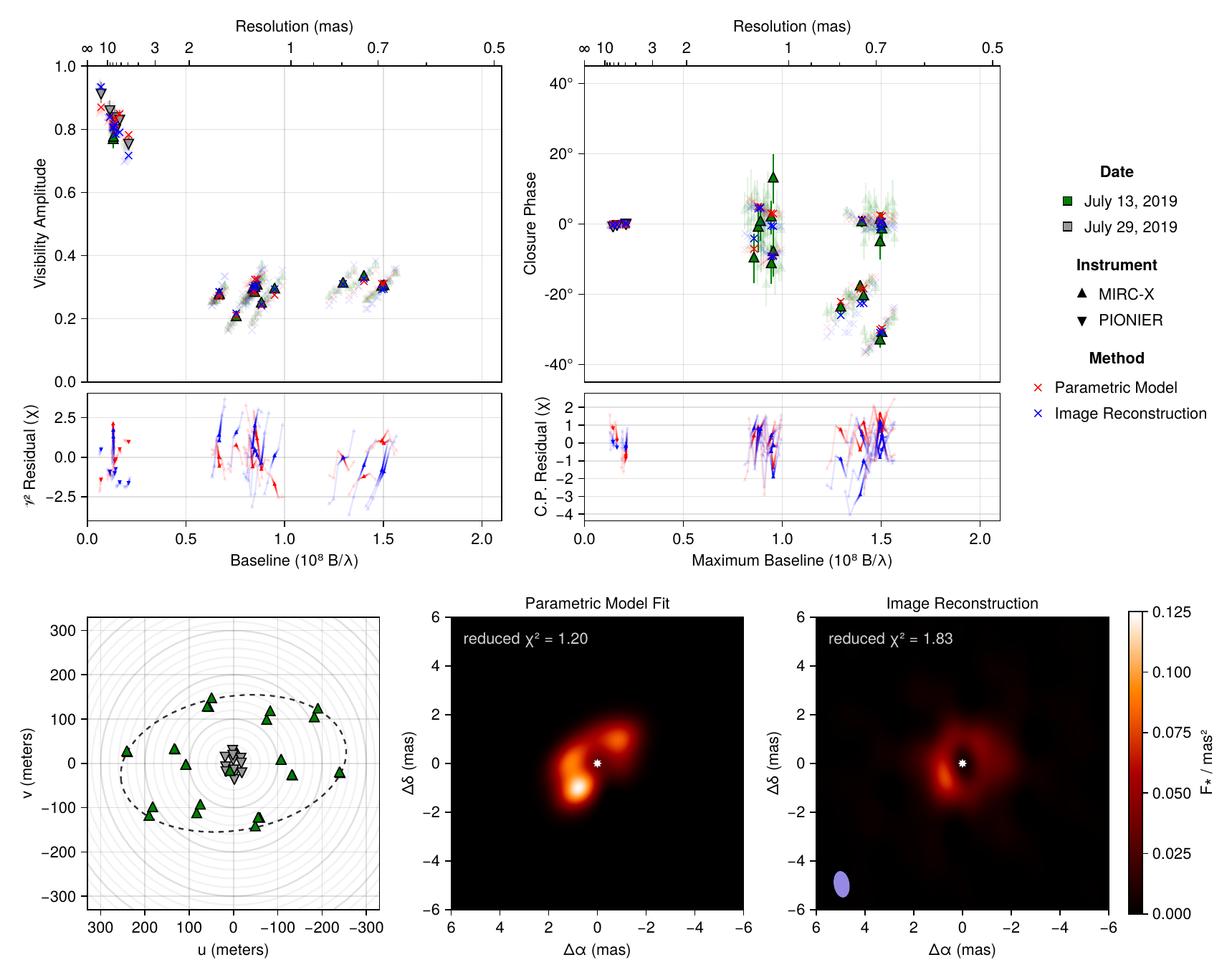}
    \caption{Summary of observations and model fits to the July 13, 2019 (\mircx) epoch data. See the caption in Fig~\ref{fig:summary_june5} for an explanation of what is depicted in the panels.}
    \label{fig:summary_july13}
\end{figure*}

\begin{figure*}
    \centering
    \includegraphics[width=\textwidth]{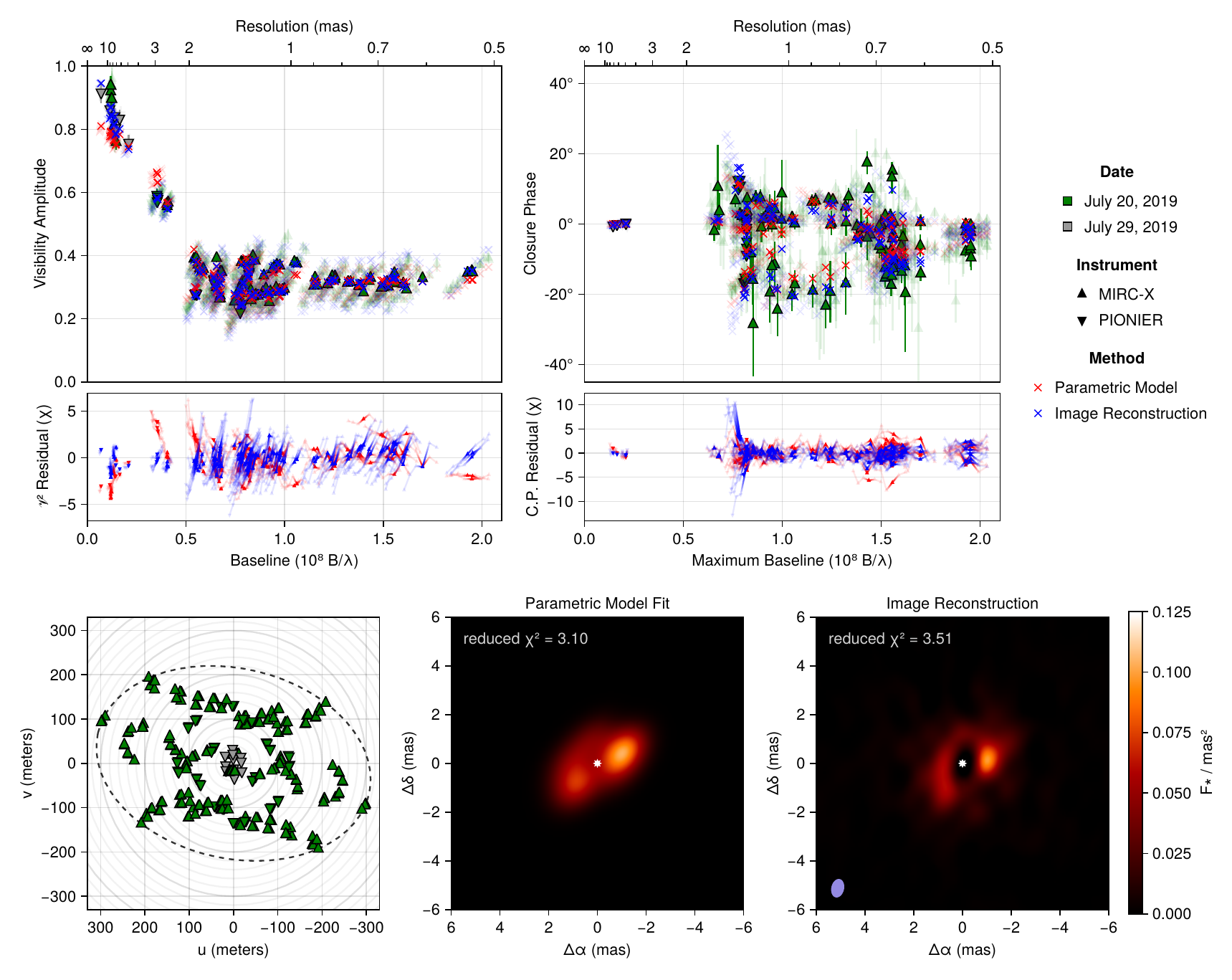}
    \caption{Summary of observations and model fits to the July 20, 2019 (\mircx and PIONIER) epoch data. See the caption in Fig~\ref{fig:summary_june5} for an explanation of what is depicted in the panels.}
    \label{fig:summary_july20}
\end{figure*}

Figures \ref{fig:summary_june5}--\ref{fig:summary_july20} demonstrate individual night fits (parametric modeling and image reconstruction) with respect to interferometric observable data. Note that while data are actually fit to square visibility measurements, the following figures display the square root of this value (a proxy for visibility amplitude) instead as this makes interpretation of the various emission components conceptually easier.

\begin{figure*}
    \centering
    \includegraphics[width=\textwidth]{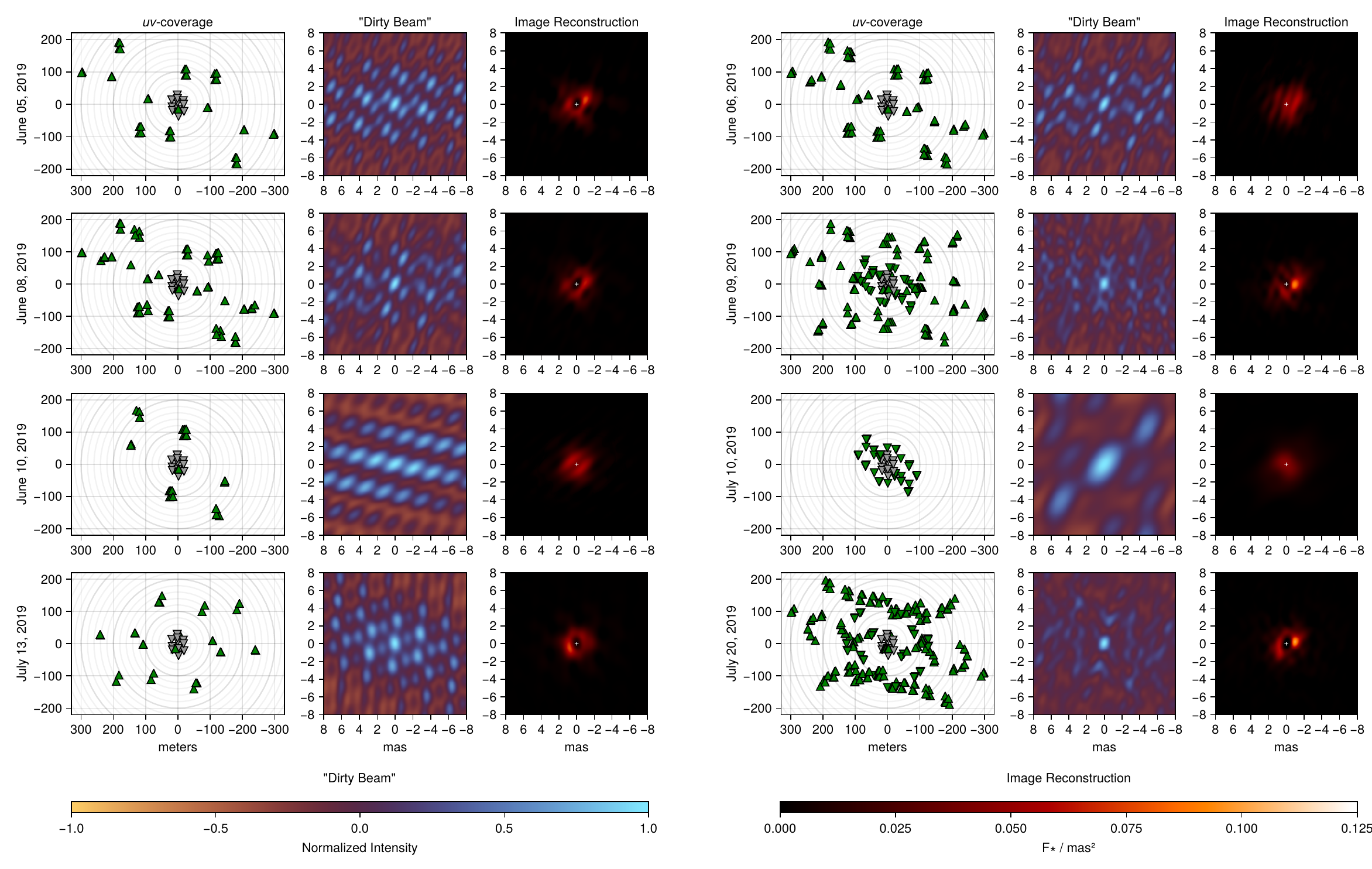}
    \caption{We remind the reader that a CLEAN algorithm was \emph{not used} in the image reconstruction procedure (owing to the impossibility of direct visibility phase measurement in the infrared through Earth's atmosphere). Nevertheless, a ``dirty beam'' map still may still provide some readers with insights related to interferometric resolution (size of the ''primary beam'', compare to the beam size in Figs. \ref{fig:summary_june5}--\ref{fig:summary_july20}) as well as a vague qualitative notion of the liable morphology of imaging artifacts.}
    \label{fig:beam_map}
\end{figure*}

Figure \ref{fig:beam_map} provides a digestible comparison of the relative $uv$-coverage amongst individual epochs, together with their associated ``dirty beam'' maps, which may be informative to readers with a radio interferometric background. We would like to use the plots in this figure to highlight 2 key takeaway messages: (1) The sparsity of long baseline observations in the June 10 (CHARA with only 4 telescopes) and July 10 (VLTI only) epochs explains the comparatively poorer quality of the recovered images and (2) The first three epochs have very similar overall $uv$-coverage, yet the image reconstructions (and analytical fits) recover a dominant spot location that moves over time. Whereas some other transient features do appear to be recovered in the analytical fits and reconstructed images, their inconsistency between epochs, coupled with the overall limitations in $uv$-coverage (including prominent ``side-lobes'' in most epochs), makes it more difficult to discriminate them from reconstruction artifacts. Future observation campaigns with higher cadence among epochs will be better equipped to comment on these features' veracity.

\section{Code Availability}

Code developed and used in this analysis for parametric modeling and image reconstruction is made available at \mbox{\url{https://gitlab.chara.gsu.edu/bensett/setterholm_et_al_2025}}. Together with the OIFITS data, accessible via OIDB \mbox{(\url{http://oidb.jmmc.fr/collection.html?id=cb5b2fd1-e23a-4c6c-9b2f-d1e23aec6c95})}, everything required to reproduce the fits and images shown in this manuscript is present.

\end{document}